\documentclass{svjour3}

\RequirePackage{fix-cm}
\smartqed  

\usepackage[T1]{fontenc} 
\usepackage[sort]{natbib}
\usepackage{amsmath,amssymb,commath, amsfonts}
\usepackage{subfig}
\usepackage{graphicx}
\usepackage{booktabs, tabularx}
\usepackage{multirow}
\usepackage{parcolumns}
\usepackage[most]{tcolorbox}

\usepackage[colorlinks]{hyperref}
\hypersetup{linkcolor=black, citecolor=black, urlcolor=black}
\newcolumntype{L}{>{\hsize=.2\hsize\raggedright\arraybackslash}X}
\newcolumntype{R}{>{\hsize=.16\hsize\raggedleft\arraybackslash}X}

\usepackage{color}
\usepackage{listings}
\usepackage{framed}

\definecolor{keyword}{RGB}{127,0,0}
\definecolor{block}{RGB}{0,127,0}
\definecolor{variable}{RGB}{0,0,127}

\begin{document}
\sloppy
\title{Smells in System User Interactive Tests}

\author{Renaud Rwemalika      \and
        Sarra Habchi          \and
        Mike Papadakis        \and 
        Yves Le Traon         \and
        Marie-Claude Brasseur
}

\institute{
        Renaud Rwemalika \at University of Luxembourg \email renaud.rwemalika@uni.lu   \and
        Sarra Habchi \at University of Luxembourg \email sarra.habchi@uni.lu           \and
        Mike Papadakis \at University of Luxembourg \email michail.papadakis@uni.lu    \and 
        Yves Le Traon \at University of Luxembourg \email yves.letraon@uni.lu          \and
        Marie-Claude Brasseur \at BGL BNP Paribas \email marie-claude.brasseur@bgl.lu 
}

\maketitle

\section*{Abstract}

Test smells are known as bad development practices that reflect poor design and implementation choices in software tests.
Over the last decade, test smells were heavily studied to measure their prevalence and impacts on test maintainability.
However, these studies focused mainly on the unit level and to this day, the work on system tests that interact with the System Under Test through a Graphical User Interface remains limited. To fill the gap, we conduct an exploratory analysis of test smells occurring in System User Interactive Tests (SUIT). 
First, based on a multi-vocal literature review, we propose a catalog of 35  SUIT-specific smells. 
Then, we conduct an empirical analysis to assess the prevalence and refactoring of these smells in 48 industrial test suites and 12 open-source projects. We show that the same type of smells tends to appear in industrial and open-source projects, but the symptoms are not addressed in the same way. Smells such as Obscure Test, Sneaky Checking, and Over Checking show symptoms in more than 70\% of the tests. 
Yet refactoring actions are much less frequent with less than 50\% of the affected tests ever undergoing refactoring. Interestingly, while refactoring actions are rare, some smells, such as Narcissistic, disappear through the removal of old symptomatic tests and the introduction of new tests not presenting such symptoms.

\section{Introduction} 

Today, Graphical User Interface (GUI) are becoming ubiquitous as a means for users to interact with a software application \citep{Myers1992, Myers1995, Brooks2009, Memon2010}.  As a consequence of the predominance of GUI-based applications, a lot of effort has been geared towards GUI test automation, notably focusing on desktop applications \citep{Nguyen2014, Advolodkin2018, Pezze2018}, web applications \citep{Mesbah2009, Biagiola2019}, mobile applications \citep{Machiry2013, Gomez2013, Mao2016, Salihu2019, Yu2019} or cross-platform applications \citep{Canny2020}.

Thanks to the maturity of test automation frameworks, both commercial and open-source projects are increasingly adopting automated test suites to ensure software quality. However, testing interactive applications remains a major challenge for software projects.  Indeed, similarly to production code, quality issues might emerge in the test code. For example, prior work showed the existence of sub-optimal design choices that lower the quality of the test code \citep{VanDeursen2001, Meszaros2007, Reichhart2007, VanRompaey2007, Bavota2015, Tufano2016, Bowes2017, Kim2020, Peruma2020} in the form of test smells.

The concept of smells was first introduced by \cite{Fowler1999}, who defined code smells as poor design and implementation choices that hinder the system maintainability. Later the concept of smells was extended to cover maintainability issues in test code~\citep{VanDeursen2001}. Test smells differ from classical code smells by covering the organization, implementation, and interactions of software tests~\citep{Moonen2008}.

In this paper, we focus on smells that appear in tests residing at the higher-levels of the test pyramid, namely System User Interactive Tests (SUITs). More specifically, a SUIT can be defined as an automatic test exercising a software application with a graphical front-end to guarantee that it meets its specification by performing a sequence of events against the GUI elements \citep{Cunha2010, Banerjee2013, Issa2012}. As such, these tests present unique characteristics in their (1) structure, (2) interaction with the System Under Test (SUT), and (3) the actors involved in their maintenance.

Indeed, interacting through the user interface, SUITs present unique characteristics that originate from their mode of interaction and synchronization with the SUT. SUITs consider the SUT as a black box, ignoring its implementation details to test higher behavioral components. Thus, SUITs present fundamental differences in the way they are crafted when compared to unit tests. For example, the smell \emph{Mystery Guest}, which is well studied in test smell literature \citep{VanDeursen2001, Bavota2015, Tufano2016, DeBleser2019, Peruma2020, Virginio2020} appears when a test class relies on an external resource file. However, in the case of SUITs, extracting away data is considered a good practice and is highly used in data-driven testing \citep{Baker2008} in order to decouple the data from the behavior specifications. Indeed, one of the functions of a SUIT being communication between different roles, namely business analysts, testers and developers, technical details are hidden away to ease communication between the stakeholders. Our study aims to understand the effects of test smells when accounting for the unique characteristics present in SUITs. Specifically, we aim to answer the following research questions:
\begin{itemize}
    \item \textbf{\textsc{RQ1}}: What are the SUIT smells studied in academic and grey literature?\\
    \textbf{Goal}: This question aims to explore and identify test  smells that are specific to SUITs. 
    Identifying and categorizing test smells that are mentioned and  discussed in the grey literature bridges the gap between research community and practitioners since the majority of grey literature is based on the practitioners views. In view of this, our intent is to produce an almost exhaustive catalog of known SUIT smells along with their definition and impact. 
    \\
    \item \textbf{\textsc{RQ2}}: How widespread are SUIT smell symptoms in SUITs?\\
    \textbf{Goal}: This question aims at assessing the prevalence of SUIT smells in industrial and open-source projects and how different smells appear in the test codebases. An additional aim in analysing this question is the analysis of how smells are viewed and managed by practitioners. Empirically answering this question provides a sanity check on the importance of the SUIT smells,  i.e., very rare smells are probably not interesting as they are not part of the current practice. 
   \\
    \item \textbf{\textsc{RQ3}}: How often do we observe refactoring actions removing SUIT smell symptoms?\\
    \textbf{Goal}: This question aims to analyse the refactoring actions performed by maintainers to remove SUIT smells. While there may be a large amount of smells present in the test codebase, practitioners may be unaware of their presence. To explore this, we identify and detect fine-grained smell refactoring operations that reflect an interest by the maintainers of the test suite. Answering this question provides evidence in the importance of the smells, i.e., they are associated with development time and effort. 
    \\
\end{itemize}

To answer these questions, we combine a multi-vocal literature review and an empirical study on a large industrial project and 12 open-source repositories. Relying on the multivocal literature review, we answer RQ1 and build a catalog of 35 SUIT smells. For 16 out of the 35 SUIT smells of this catalog we propose an automated approach for detecting their diffusion and refactoring. Leveraging this approach, we evaluate the prevalence of SUIT smells and their removal in more than two million tests from our dataset composed of industrial and open-source projects. The main contributions of this study are:

\begin{itemize}
    \item We introduce the first catalog of SUIT-specific smells building on the knowledge of both researchers and practitioners. Alongside with it, we propose an open-source automated tool\footnote{Available at https://github.com/UL-SnT-Serval/ikora-evolution} to detect symptoms and refactorings for 16 SUIT smells.
    \item We show that the same type of smells tend to appear in industrial and open-source projects, but the symptoms are not addressed in the same way.
    The smells \emph{Hard Coded Values} (85\% to 90\% of tests), \emph{Over Checking} (75\% to 80\% of tests), and \emph{Sneaky Checking} (around 70\% of tests) are the most prevalent SUIT smells in both types of projects.
    \item Our empirical results suggest that less than half of SUITs ever experience smell refactorings. 
    \emph{Missing Assertion} is a unique exception with up to 90\% of the symptomatic tests refactored. Interestingly, while refactoring actions are rare, SUIT smells like \emph{Narcissistic} and \emph{Middle Man} still disappear from the codebase as a side effect of unintended maintenance and refactoring operations.
\end{itemize}

\section{Background}

This section presents Keyword-Driven Testing (KDT), a popular paradigm to create SUITs, along with an illustrative example and a description of its supporting framework, Robot Framework\footnote{https://robotframework.org/}.

\subsection{Keyword-Driven Testing}
\label{sec:background-keyword-driven-testing}

KDT is a software testing technique using \emph{Keywords} where each \emph{Keyword} describes a set of actions that are required to perform a specific step. 

Through the use of \emph{Keywords}, KDT aims at separating test design from the technical implementation of tests, thus, limiting exposure to unnecessary details and avoiding duplication. KDT advocates that this separation of concerns makes tests easier to write and more maintainable. On top of that, the separation enables experts from different fields and backgrounds to work together, at different levels of abstraction~\citep{tang2008}. For these reasons, KDT is an ideal candidate to write SUITs.

\cite{Humble2010} formalize a hierarchy under three distinctive layers: (1) the acceptance criteria which describes the functional behavior, usually written in a form close to natural language; (2) the application driver layer that contains the underlying implementation of test using the vocabulary from the application domain; and (3) the application driver layer that understands how to interact with the SUT and is expressed in the domain language of the driver that is used to communicate with the SUT. 

Figure~\ref{fig:robot-script} shows an example of a KDT test adapted from the official Robot Framework documentation. The test \texttt{A user logs in with his username and password} (line 6) is responsible for validating the correct behavior of the login form in an imaginary SUT by interacting with its user interface.

As can be seen from the listing, most parts of this fully automated test are written in plain English. This enables the unobstructed collaboration in the creation and analysis of the tests between different experts. To further extend the reuse of \emph{Keywords}, they can have \emph{Arguments}. For instance, the \emph{Library Keyword} \texttt{Open browser} (line 21) takes two arguments, \texttt{\$\{LOGIN URL\}} and \texttt{\$\{BROWSER\}}.

Lines 6--8 present the \texttt{test steps}, which represent the test acceptance criteria. In KDT, each step is implemented as a \emph{Keyword}. In turn, these \emph{Keywords} are defined in their respective definition blocks between lines 12 and 41. The actions composing the body of a \emph{Keyword} are subsequent \emph{Keyword} calls.

A the end of the call tree, we find \emph{Library Keywords} performing concrete actions such as interacting with the SUT or managing the test execution. For example, in Figure~\ref{fig:robot-script}, line 2 shows that the script is using the web driver automation library \emph{Selenium} to interact with the SUT. This library allows the test script to perform concrete actions such as navigating through the website using the \texttt{Go To} \emph{Library Keywords} at line 26 and interact with elements present on a web page with library keywords such as \texttt{Input Text} (lines 31 and 35) or \texttt{Click Button} (line 38). Furthermore, \emph{Library Keywords} can be used to perform assertions such as \texttt{Title Should Be} at line 41 ensure that the title (title element in the header of the DOM) is of a specific value.

Thus, KDT makes it easy to follow the three-layer architecture described by \cite{Humble2010}. In this model, the \emph{Test Case} and its steps are the acceptance criteria, the intermediate \emph{User Keywords} are the implementation layer and finally, the \emph{Library Keywords} interacting with the system under test represent the application driver layer.

\begin{figure}
\centering
\caption{Example of Robot Framework test}
\label{fig:robot-script}
\begin{minipage}{0.8\linewidth}
\begin{lstlisting}
    <@\textcolor{block}{*** Settings ***}@>
    Test Teardown     Close All Browsers
    Library           Selenium2Library
    
    <@\textcolor{block}{*** Test Cases ***}@>
    <@\textcolor{keyword}{A user logs in with his username and password}@>
        Given browser is opened to login page
        When user "demo" logs in with password "mode"
        Then welcome page should be open
    
    <@\textcolor{block}{*** Keywords ***}@>
    <@\textcolor{keyword}{Browser is opened to login page}@>
        Open browser to login page
    
    <@\textcolor{keyword}{User "}@><@\textcolor{variable}{\$\{username\}}@><@\textcolor{keyword}{" logs in with password "}@><@\textcolor{variable}{\$\{password\}}@><@\textcolor{keyword}{"}@>
        Input username    <@\textcolor{variable}{\$\{username\}}@>
        Input password    <@\textcolor{variable}{\$\{password\}}@>
        Submit credentials
    
    <@\textcolor{keyword}{Open Browser To Login Page}@>
        Open Browser    <@\textcolor{variable}{\$\{LOGIN URL\}}@>    <@\textcolor{variable}{\$\{BROWSER\}}@>
        Maximize Browser Window
        Title Should Be    Login Page        
    
    <@\textcolor{keyword}{Go To Login Page}@>
        Go To    <@\textcolor{variable}{\$\{LOGIN URL\}}@>
        Login Page Should Be Open
    
    <@\textcolor{keyword}{Input Username}@>
        [Arguments]    <@\textcolor{variable}{\$\{username\}}@>
        Input Text    username_id    <@\textcolor{variable}{\$\{username\}}@>
    
    <@\textcolor{keyword}{Input Password}@>
        [Arguments]    <@\textcolor{variable}{\$\{password\}}@>
        Input Text    password_id    <@\textcolor{variable}{\$\{password\}}@>
    
    <@\textcolor{keyword}{Submit Credentials}@>
        Click Button    validate_id
        
    <@\textcolor{keyword}{Welcome Page Should Be Open}@>
        Title Should Be    Welcome Page
    
    <@\textcolor{block}{*** Variables ***}@>
        <@\textcolor{variable}{\$\{SERVER\}}@>           localhost:7272
        <@\textcolor{variable}{\$\{BROWSER\}}@>          Chrome        
        <@\textcolor{variable}{\$\{LOGIN URL\}}@>        http://<@\textcolor{variable}{\$\{SERVER\}}@>
\end{lstlisting}
\end{minipage}
\end{figure}

\subsection{Robot Framework}
\label{sec:background-robot-framework}

Robot Framework is an open-source KDT framework that was originally developed by Nokia Networks and is widely adopted by companies. Robot framework relies on human-readable keywords that make the code easy to understand by non-technical people. Indeed, as shown in Section~\ref{sec:background-keyword-driven-testing},  the acceptance criteria of the test \texttt{A user logs in with his username and password} in Figure~\ref{fig:robot-script} (lines 6-9), written using the Robot Framework syntax, is entirely expressed using natural language.

One of the main advantages of Robot Framework is its high modularity. The framework is platform-agnostic and thanks to its driver plugin architecture, the core framework does not require any knowledge of the SUT. Indeed, libraries accomplishing the concrete actions are created either by the Robot Framework core team and integrated in the language, developed by third parties and distributed as library plugins, or developed directly by the test team in which case the library code can be maintained alongside the test code. Thus, the concrete implementation of an action in a Robot Framework test script is hidden from the test. On the down side, while good practices promote such isolation and separation of concerns, the introduction of suboptimal decision can hinder the readability of the test code, alter interactions with the SUT, or just make it harder to adapt the test codebase to SUT evolution.

\begin{figure}
\centering
\includegraphics[width=0.7\linewidth]{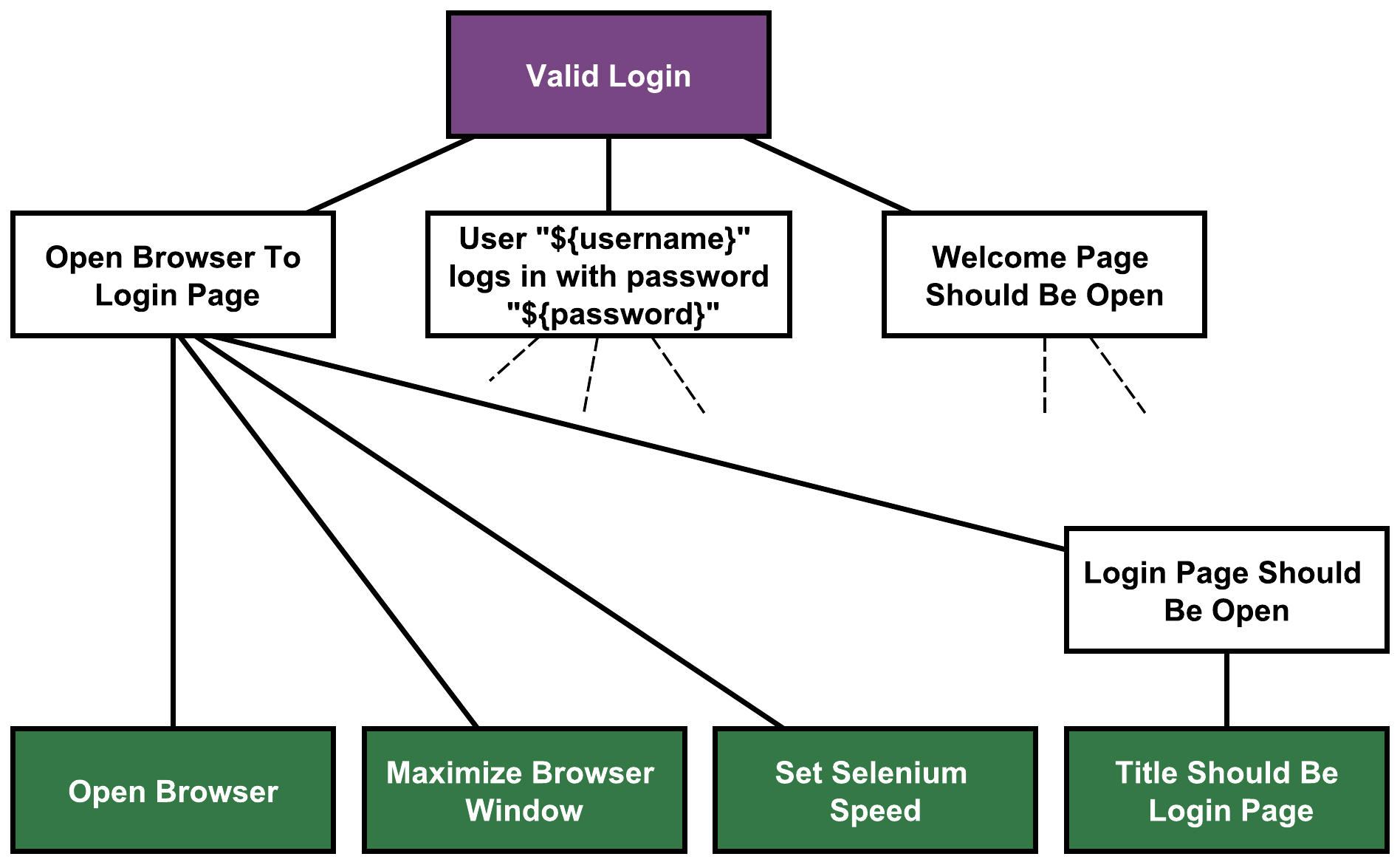}
\caption{Tree representation of a Robot Framework test~\citep{Rwemalika2019}.}  
\label{fig:robot-tree}
\end{figure}

A Robot Framework script can be represented using a tree structure, \emph{i.e.} a rooted, ordered, directed, acyclic graph. Figure~\ref{fig:robot-tree} shows the tree representation of the test presented in Listing~\ref{fig:robot-script}. The nodes of the tree represent an entity from the Robot Framework language (\emph{Test Case}, \emph{Keyword} or \emph{Variable}) and the edges represent \emph{Keyword} calls. At the root of the tree, we find the \emph{Test Case} (purple in the figure). Each \texttt{test sep} composing the acceptance criteria is represented by a call from the \emph{Test Case} to a \emph{Keyword}. The intermediate nodes (white in the figure) represent \emph{User Keywords}. \emph{User Keywords} are calling subsequent \emph{Keywords} which can be either other \emph{User Keywords} or \emph{Library Keywords}. Finally, the leaves nodes are the \emph{Library Keywords} (in green in the figure). Because they perform concrete actions that are hidden from the test script, they do appear in the graph as leaf nodes. When building the graph representation of the tests, we further annotate the \emph{Library Keyword} with a category defining which type of action they perform \citep{Rwemalika2019}, namely: INTERACTION, ASSERTION, CONTROLFLOW, GETTER, LOGGING and SYNC.

Building on this model we rely on graph theory algorithms to traverse all the nodes of a test and extract properties of the tests and the symptoms presented in Section~\ref{sec:results-smells-catalog}.

\section{Experimental Design}
\label{sec:methodology}

\subsection{RQ1: Identification of SUIT Smells}
\label{sec:experience-design-smells-collection}

\begin{figure}
    \centering
    \includegraphics[width=\linewidth]{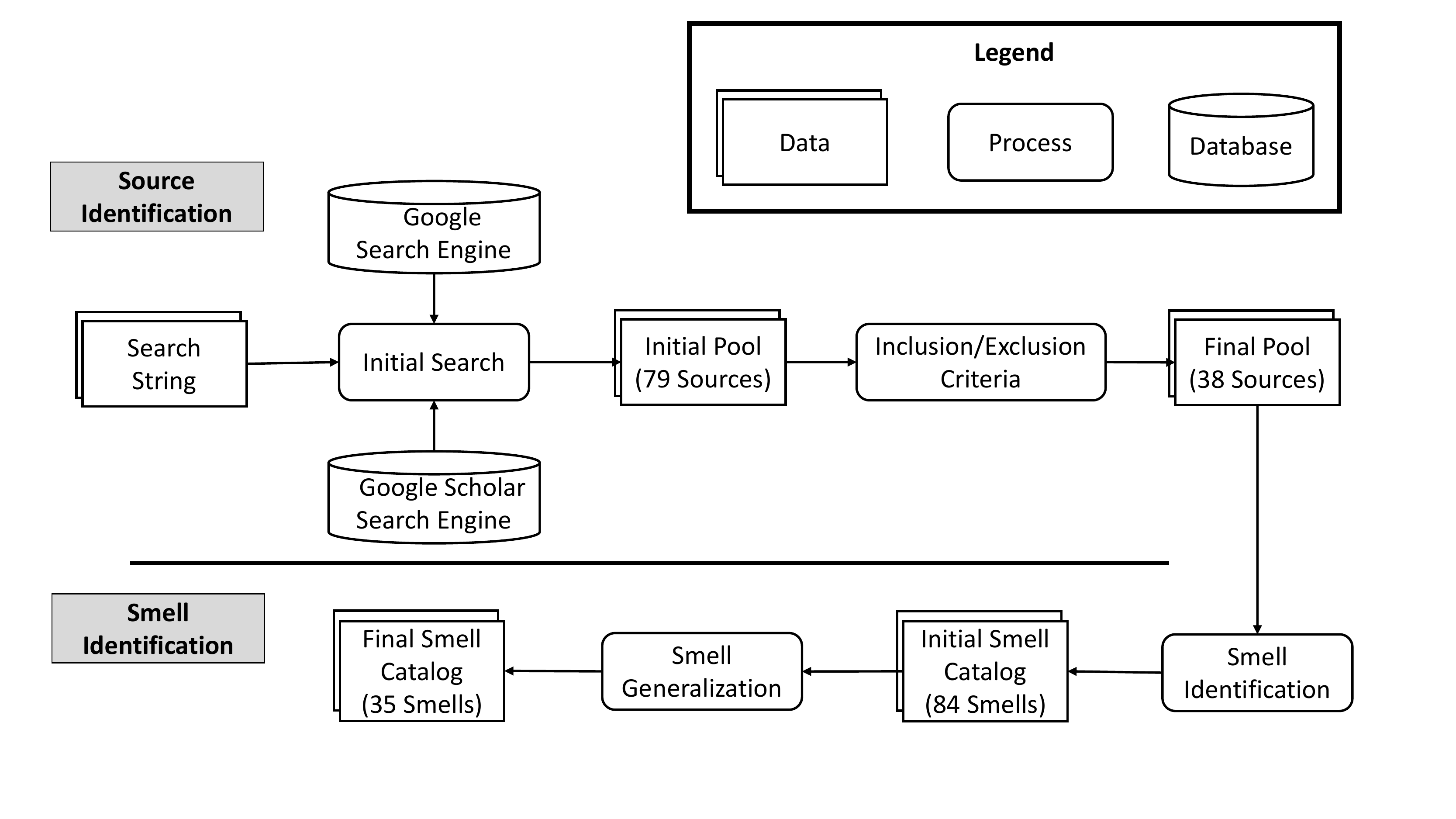}
    \caption{Overview of the process to establish the SUIT smell catalog.}  
    \label{fig:smell-catalog-process}
    \end{figure}

To perform a study on the impact of smells in SUITs, we first need to build a reliable and representative catalog of test smells. To this end, we start our investigation by collecting SUIT smells presented in both academic and grey literature. 
We classify in the academic or white literature papers that are published in peer reviewed conferences or journals. On the other hand, we classify as grey literature white-papers, magazines, online blog-post, question-answers sites, survey results and technical reports following the methodology presented in \cite{Ricca2021}. Indeed, the grey literature constitutes a rich source of documents where practitioners share their experiences, propose guidelines or even ask and answer questions. Thus, we conduct a multivocal literature review following the steps depicted by \cite{Garousi2018}. 
Figure~\ref{fig:smell-catalog-process} summarizes our adoption of these steps in the process of building a catalog of SUIT smells.

\paragraph{\textbf{Initial Search.}}

To mine the academic sources we rely on the Google Scholar Search engine, whereas for the grey literature we rely on the Google Search engine. These two search engines are known to subsume other databases and repositories hence we limit our investigation to them. We compile a list of search terms such as ``acceptance test'', ``test automation'', ``end-to-end test'', ``system test'' and ``behavior test'' to define the type of tests that we are targeting and combine those strings with each of the following: ``smell'', ``symptom'', ``anti pattern'' and ``bad practice''. This allows to generate the following search string for the engines: \emph{software and ("acceptance test" OR "test automation" OR "end-to-end test" OR "system test" OR "functional test") AND ("smell" OR "symptom" OR "anti pattern" OR "bad practice")}. To validate the search string we run it and make sure that relevant documents known \emph{a priori} appear in the results \citep{Kitchenham2007, Ricca2021}. 
Then, we proceed to a systematic review of all the sources in order to identify test smells. However, because the search string returns about 1,370,000 results in the Google Search Engine and about 12,100 results in the Google Scholar Search Engine, we restrict our analysis to an initial pool composed of the first 200 results in each engine. From those we read the title and abstract available to decide whether or not to further analyze the articles. This leads to a initial list composed of 47 sources in the grey literature and 32 sources in the academic literature for a total of 79 sources.

\paragraph{\textbf{Inclusion/Exclusion Criteria.}}

To be eligible for our study, a smell must address the code of a SUITs. However, the terms adopted in both industry and academia might not refer to how the test interact with the application, but rather to what is the intent of the test (\emph{e.g.} acceptance testing) or its scope (\emph{e.g.} system test). This leads to a series of sources discussing tests that are not SUITs and are not compatible with them, typically, white-box tests. Furthermore, some sources address testing issues that are not related to the codebase such as organizational smells, which are outside the scope of this study. Indeed, some smells do not target the test code but testers behavior \emph{e.g.}, \emph{Making Intermittent Bugs Low Priority} \citep{StackExchange2017} for which bugs making the functional test suite fail intermittently at low intervals are ignored. By excluding these irrelevant sources, we reduce our list to 32 sources from the grey literature and 6 sources in the academic literature for a total of 38 sources. The drastic drop in the academic literature is due to the large number of studies targeting unit tests instead of SUITs.
The selection process is depicted in the summary sheet\footnote{Available as \href{https://docs.google.com/spreadsheets/d/e/2PACX-1vQ78jmOjU3qTOlGzwCSkidJOliPKNDQhmuOxSsfTaRqFVjmFP41JUbYQeupqU_lGCK6L4EpQ3FHNGhU/pubhtml}{a spreadsheet}}.

\paragraph{\textbf{Smell Identification.}}

Relying on this final pool of sources, where each source contains at least one SUIT smell, we proceed to the smell identification where we filter out all the smells that are not suitable for our analysis. Indeed, some smells are not applicable to the framework under study \emph{e.g.}, \emph{Dependence on Record and Playback} \citep{StackExchange2017} in which tests are created using Record and Playback, a feature that is not supported by Robot Framework. Other smells are too technology specific \emph{e.g.}, not using the page object pattern in Selenium tests \citep{Advolodkin2018}. The outcome of this process is an initial smell catalog of 84 unique SUIT smells for which we can derive a metric by analyzing the code representing the symptom observed.

\paragraph{\textbf{Smell Generalization.}}
Some SUIT smells gathered in the previous step exhibit large overlaps; thus, we proceed to a smell generalization step. For instance, the smell \emph{Enter Enter Click Click} \citep{Buwalda2015} is grouped with \emph{Comments and documentation instead of abstraction} \citep{Klarck2014}, since both smells target the presence of low level actions in the acceptance criteria but in the latter a specific emphasis is put on the documentation aspect. Hence, we consider those two smells to present similar symptoms and effects and we group them in one smell named \emph{Lack of encapsulation} \citep{Chen2012, Klarck2014, Buwalda2015, England2016, Renaudin2016}. Finally, we observe that some test smells are subsuming others. This is the case for the smells \emph{Hardcoded Values} and \emph{Using condition in test logic} that are both subsumed by the definition of \emph{Obscure test} \citep{Hauptmann2013, Gawinecki2016, Siminiuc2019}. In this example we only keep the latter.  Finally, from this list, we filter out any test smell that would require the test to be executed in order to be observed. As a result, the outcome of this step is a list of 35 SUIT smells\footnote{The complete catalog is available at https://github.com/UL-SnT-Serval/suit-smell-catalog} that can be detected statically.

Finally, conducting our study using Robot Framework, we exclude four test smells that cannot be observed in this specific language and its associated framework. Furthermore, we omit 15 test smells because, despite our best efforts, no automated metric avoiding false positive could be constructed by analyzing the test code. Hence, in this work, we present a list of 16 SUIT smells. A comprehensive description of each smell is presented in Section~\ref{sec:results-smells-collection}.

\subsection{Dataset}
\label{sec:data-collection}
To answer RQ2 and RQ3, we conduct a case study on 13 test suites written in Robot Framework. We establish two sets of projects: the first set is composed of 48 repositories from our partner site, BGL BNP Paribas and the second set of projects is composed of 12 open-source projects mined from Github. Table~\ref{tab:projects} summarizes the overall properties of these projects. In the following we describe the projects and present the collection process. 

\subsubsection{Industrial project:}
We leverage the codebase of our industrial partner, BGL BNP Paribas. The test suite is maintained by a dedicated quality assurance team which role is to ensure the compliance to the requirements of the products deployed to production. Thus, the team tests desktop, web and mobile applications that are depending on services developed following a service-oriented architecture (SOA). The goal of the team is to assessing compliance to functional requirements, applications are tested in a black box fashion through their user interface.

Historically, the team relied on manual testing to perform its tasks. However, with the release cycles becoming shorter and the number of tests increasing, the execution burden on the team became unmaintainable. Thus, in the end of 2016, the team started migrating from an manual execution of the tests to the execution of automated tests written in Robot Framework. Robot Framework was chosen because of the ease it provides to develop tests targeting application written in different technologies requiring different mode of interactions.

Today, the test suite consists of 559 tests stored across 48 repositories on an on premise GitLab instance. While some repositories are defining \emph{Test Cases}, others are used as resources of common \emph{Keywords} where a series of generic \emph{Keywords} specific to the BGL BNP Paribas architecture have been created to help with the development effort as well as to avoid code duplication. A typical example is the login to the ecosystem which is common to a large amount of services, and consequently can be mutualized. Hence, in this study, we merge all the repositories to count them as one project, which explains the larger number of \emph{User Keywords} observed in Table~\ref{tab:projects} for the project \emph{bgl}.

\subsubsection{Open-source projects:} 
To collect the open-source test suites, we use the Github Search API to mine repositories containing suitable test suites, \emph{i.e} Robot Framework test suites. From the results of this first step, we filter out toy projects and tutorials, projects using Robot Framework for its robot process automation capabilities in production and libraries extending the capabilities of Robot Framework. Following this approach, we gather 23 repositories from Github. Finally, to ensure that maintenance was performed on the test suite itself, we analyze the number of commit involving the Robot Framework test suite for each project, \emph{i.e.} SUIT modifying commit, and discard any project with less than 100 SUIT modifying commits. This process yields a total of 12 repositories presented in Table~\ref{tab:projects}.

The data collection process leads to a total of 2,884,383 SUITs analyzed across all the versions of all the projects where 2,742,271 originate from the open-source projects and 142,112 from the industrial projects gathered at BGL BNP Paribas.

\begin{table}
\centering

\begin{tabularx}{\textwidth}{LRRRR}

\hline
\textbf{Name} & \textbf{LoC} & \textbf{\#Commits} & \textbf{\#TestCases} & \textbf{\#Keywords} \\
\hline

\scriptsize{\textbf{bgl}} & \scriptsize{57030} & \scriptsize{309} & \scriptsize{559} & \scriptsize{5000} \\

\hline

\scriptsize{\textbf{apinf}} & \scriptsize{1068} & \scriptsize{345} & \scriptsize{74} & \scriptsize{138} \\
\scriptsize{\textbf{bikalims}} & \scriptsize{4446} & \scriptsize{1279} & \scriptsize{76} & \scriptsize{172} \\
\scriptsize{\textbf{collective-cover}} & \scriptsize{1317} & \scriptsize{943} & \scriptsize{24} & \scriptsize{31} \\
\scriptsize{\textbf{cumulus-ci}} & \scriptsize{1599} & \scriptsize{1429} & \scriptsize{132} & \scriptsize{34} \\
\scriptsize{\textbf{harbor}} & \scriptsize{7270} & \scriptsize{3131} & \scriptsize{246} & \scriptsize{472} \\
\scriptsize{\textbf{ifs}} & \scriptsize{8466} & \scriptsize{5606} & \scriptsize{494} & \scriptsize{605} \\
\scriptsize{\textbf{mms-alfresco}} & \scriptsize{1797} & \scriptsize{504} & \scriptsize{156} & \scriptsize{8} \\
\scriptsize{\textbf{mystamps}} & \scriptsize{2361} & \scriptsize{353} & \scriptsize{212} & \scriptsize{111} \\
\scriptsize{\textbf{ozone}} & \scriptsize{2636} & \scriptsize{271} & \scriptsize{230} & \scriptsize{90} \\
\scriptsize{\textbf{plone}} & \scriptsize{1046} & \scriptsize{211} & \scriptsize{59} & \scriptsize{163} \\
\scriptsize{\textbf{plone-intranet}} & \scriptsize{2225} & \scriptsize{1759} & \scriptsize{213} & \scriptsize{134} \\
\scriptsize{\textbf{rspamd}} & \scriptsize{3397} & \scriptsize{685} & \scriptsize{429} & \scriptsize{151} \\

\hline

\end{tabularx}

\caption{Metrics for the 13 projects under study. \emph{LoC} is the number of lines of Robot Framework Code, \emph{\#Commits} is the number of commits implicating Robot Framework code, and \emph{\#TestCases} and \emph{\#Keywords} are the number of test cases and user keywords respectively in the last commit of the projects.}
\label{tab:projects}

\end{table}

These 12 projects cover various technology stacks. For instance, BGL BNP Paribas developed internal services using Java Swing which rely on either the API exposed by the Swing development or rely on computer vision, through the use of the Sikuli\footnote{http://sikulix.com/} library. BGL BNP Paribas also use the Robot Framework to test its mobile client application relying on the Appium\footnote{https://appium.io/} Library. Finally, the majority of the test suites in both industrial and open-source projects from this dataset target web applications. The SUITs interact with such applications (1) through their webpages by interacting with the DOM, (2) through their APIs or (3) directly accessing the state of the database after an operation. Note that some operations generate emails or reports, thus, some tests rely on operating system interaction to check that an email was properly sent through a mail client or assess the existence and parses the content of a generated PDF documents.

To conclude this section, we observe that the dataset is composed of projects covering various domains, technology stacks, sizes and modes of development. We believe that this diversity allows to avoid biases observed in one type of projects and improves the generalization of the observations.

\subsection{RQ2: SUIT Smells Distribution}
\label{sec:smell-detection}

For each of the SUIT smells that we gathered, we compute a metric to automatically measure a smelliness score for the affected test. We rely on heuristics to construct these metrics. To this end, we apply the high-level investigation mechanism framed by \cite{Marinescu2004} called ``detection strategy''. As defined by the authors, a ``detection strategy'' is \emph{a generic mechanism for analyzing a source code model using metrics}. The detection strategies are formulated in a series of steps: (1) Break-down informal rules into symptoms that can be captured by a single metric; (2) Select a proper set of metrics to quantify each of the symptoms; (3) Define thresholds that classify a test as smelly or not; (4) Use operators to correlate the symptoms leading to the final rule for detecting the smells.

Note that the step 3 of this approach describes the definition of a threshold. Even though different approaches have been proposed, \emph{e.g.} using semantical properties and statistical distributions \citep{Marinescu2004} or using Bayesian belief networks \citep{Khomh2009}, defining a good threshold remains a hard and error prone task. Thus, more recently, researchers have been trying to avoid this limitation by using machine learning to directly learn what a smell looks like avoiding altogether the use a threshold \citep{ArcelliFontana2016}. In this work, in order to avoid any bias generated by a binary classification, we do not apply any empirical threshold (with the exception of one smell, \emph{Long Test Steps}) but focus our analysis on the observation of the symptoms associated with SUIT smells. Hence, for each test we attribute a metric that represents the number of symptoms that are observed in a test. Furthermore, we propose a density metric, which normalizes the number of symptoms observed over the worst case scenario for a given test. Indeed, our goal is to analyze how the symptoms are treated by the maintainers of the test codebase and not to classify tests as smelly or not.

To extract code metrics, we rely on a parsing engine presented in \cite{Rwemalika2019}. The parser generates for each test a call tree, where the root is the entry point of the test and the leaf nodes are the actions generated on the system under test, or the application driver layer if we refer to the three-layer architecture described by \cite{Humble2010}. Each leaf node is annotated with information such as the type of action (assertion, getter, event, etc.) and each variable can be inferred a type based on the \emph{Library Keywords} calling it. 
To create the metrics, we rely on heuristics exhibiting the symptom associated with the smell. While building the heuristics, we focus on a high precision, even though this might lead to low recall. 
For instance, for the \emph{Over-checking}, we simply get the number of assertions over the total number of nodes in the application driver layer, therefore, missing information about the position and of the assertions. Hence, we might end up missing tests exhibiting the smell when the assertions are performed prematurely, checking every action setting up the environment. However, when the ratio is high, it is a clear indication that too many assertions are made, making the intent of the test less clear.

Finally, note that for one of the SUIT smells, namely \emph{Using Personal Pronoun}, relying on code metrics alone does not provide sufficient information. Hence, we rely on textual metrics to detect the symptom. More specifically, we use natural language processing to extract information about the name of some \emph{Keywords} and use text tagging to define whether or not the subject of the \emph{Keywords} is a first person pronoun or not. All the results for this research question are presented in Section~\ref{sec:results-smells-diffusion}.

When computing the metrics, we introduce two types of metrics: a count metric $S$ and a density metric $D$. The count metric $S$ counts the number of instances of a symptom observed in a test. On the other hand, the density metric $D$ provides an indication of the number of instances over a maximum value that could have appeared in the test. Thus, the density metrics vary between 0 and 1, 0 indicating the absence of any of the symptoms and 1 indicates that for each of the possible location for the symptom to be present, it appeared in the test. Note that we focus on metrics with a high precision at the detriment of the recall. Indeed, some of the metrics only cover one form of symptoms leading to a smell but lead to low rate of false positive.

\subsection{RQ3: SUIT Smells Refactoring}
\label{sec:methodology-smell-evolution}

Refactoring can be defined as a technique to improve the design of a system and enable its evolution \citep{Fowler1999}. Relying on this definition, to answer our third research question, we conduct an analysis of the refactoring actions during the maintenance of the test suites. To do so, we collect every pair of subsequent versions and identify refactoring changes occurring in the test codebase. A key component in this type of analysis is to properly identify what constitutes a valid refactoring action. Indeed, just observing a decreases in smell metrics does not imply that a smell has been specifically addressed. Other actions such as changing the scope of a test can, as a side effect, remove symptoms of a SUIT smells without specifically targeting it. Thus, following the line of work present in the literature to identify refactoring activity \citep{Tsantalis2013, Silva2017}, we address this limitation by using heuristics based on static test code analysis. More specifically, we use the fine-grained change algorithm presented in \cite{Rwemalika2019} to extract instances of refactoring patterns. These patterns are derived from the definition of the symptoms present in the literature.

Consequently, for each SUIT smell $s$ we store the set of nodes $N$ that are exhibiting a symptom of the SUIT smell and the refactoring actions $A_{s,N}$ that when performed on one or more elements of $N$ removes the symptom. We obtain a series of tuples $<s, N, a_{s,N}>$ for each pair of subsequent versions which describes the fine-grained refactoring actions that were performed on one or more nodes to fix a specific SUIT smell symptom, \emph{i.e.} the refactoring actions. 

To generate this list, first we compile the set of potential actions $A_{s,N}$ that can fix a symptom of a SUIT smell $s$ given a set of nodes $N$ (the complete list is presented in Section~\ref{sec:results-smells-catalog}). Then, for each tuple $<N, a_{N}>$ provided by the fine-grained change extraction tool and the knowledge of which nodes present a symptom in the previous version, we check if $a_{N} \in A_{s,N}$. If it is the case, then the fine-grained change $<N, a_{N}>$ is considered to ba a refactoring action and added to the list of tuples $<s, N, a_{s,N}>$.


To assess to which extent practitioners refactor SUIT smell symptoms, we count the number of refactoring actions across all the versions, $\abs{a_{s,N}}$. Thus, the higher this number, the more time developers spend in removing the symptoms. However, because some symptoms might appear much more often than others, this number might be skewed towards symptoms that are frequent. To alleviate this limitation, we also present the number of refactoring actions over the total number of symptoms that could have been refactored by this action: $\abs{a_{s,N}} / \abs{n_{s}}$ where $\abs{n_{s}}$ is the number of nodes $n$ exhibiting a symptom $s$ across all versions. Thus, if a symptom is not removed during a multitude of commits, the metric decreases with each version where the symptom is not addressed increasing the value on the denominator ($\abs{n_{s}}$) and the numerator ($\abs{a_{s,N}}$) remaining unchanged. The results of this analysis are presented in Section~\ref{sec:results-smell-refactoring}.

\begin{table}
\centering

\begin{tabularx}{\textwidth}{>{\raggedright}p{1in}>{\raggedright}p{0.5in}>{\raggedright}p{2.9in}}

\hline
\textbf{Name} & \textbf{Notation} & \textbf{Description} \tabularnewline
\hline

\scriptsize{\# Symptoms} & \scriptsize{$S_s$} & \scriptsize{The number of symptoms for a smell $s$ counted in a single test.} \tabularnewline \noalign{\vspace{4pt}}
\scriptsize{\% Symptoms} & \scriptsize{$D_s$} & \scriptsize{The number of symptoms for a smell $s$ counted in a single test divided by the number of location they could have appeared in that test.} \tabularnewline \noalign{\vspace{4pt}}
\scriptsize{\# Refactoring} & \scriptsize{$\abs{R_{s}}$} & \scriptsize{The total number of action refactoring nodes exhibiting a symptom $R_{s}$ for a smell $s$ counted across all tests in all versions.} \tabularnewline \noalign{\vspace{4pt}}
\scriptsize{\% Refactoring} & \scriptsize{-} & \scriptsize{The proportion of symptomatic tests undergoing at least one refactoring action through their lifespan.} \tabularnewline \noalign{\vspace{4pt}}
\hline
\end{tabularx}

\caption{Definitions of the metrics used in the study.}
\label{tab:metrics}

\end{table}

\section{Results}
\label{sec:results}

\subsection{RQ1: SUIT Smells Identification}
\label{sec:results-smells-collection}

\begin{figure}
\centering
\subfloat[Source]{ \includegraphics[width=0.45\linewidth]{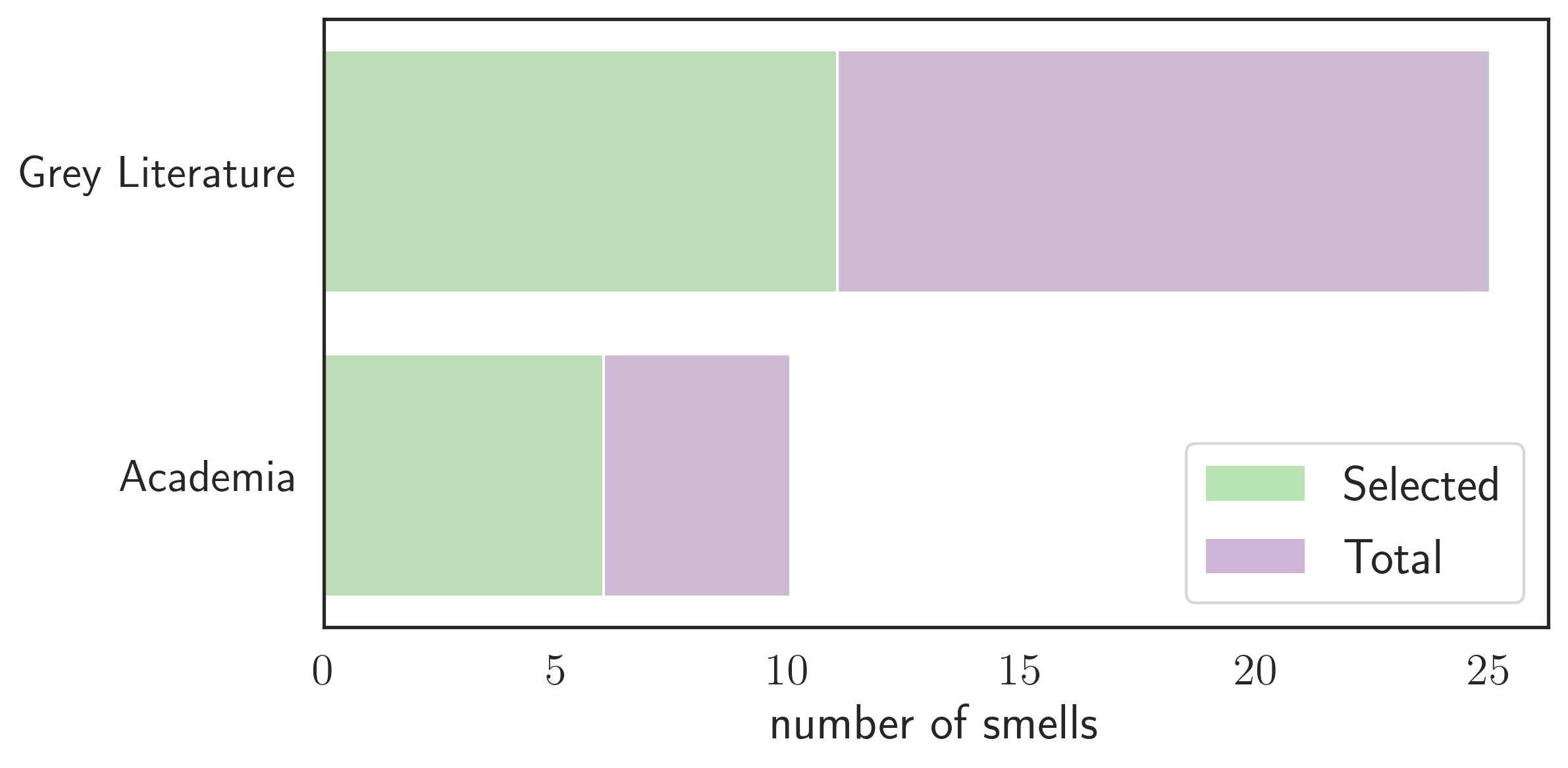}\label{fig:smell-sources}}
\subfloat[Issues]{\includegraphics[width=0.45\linewidth]{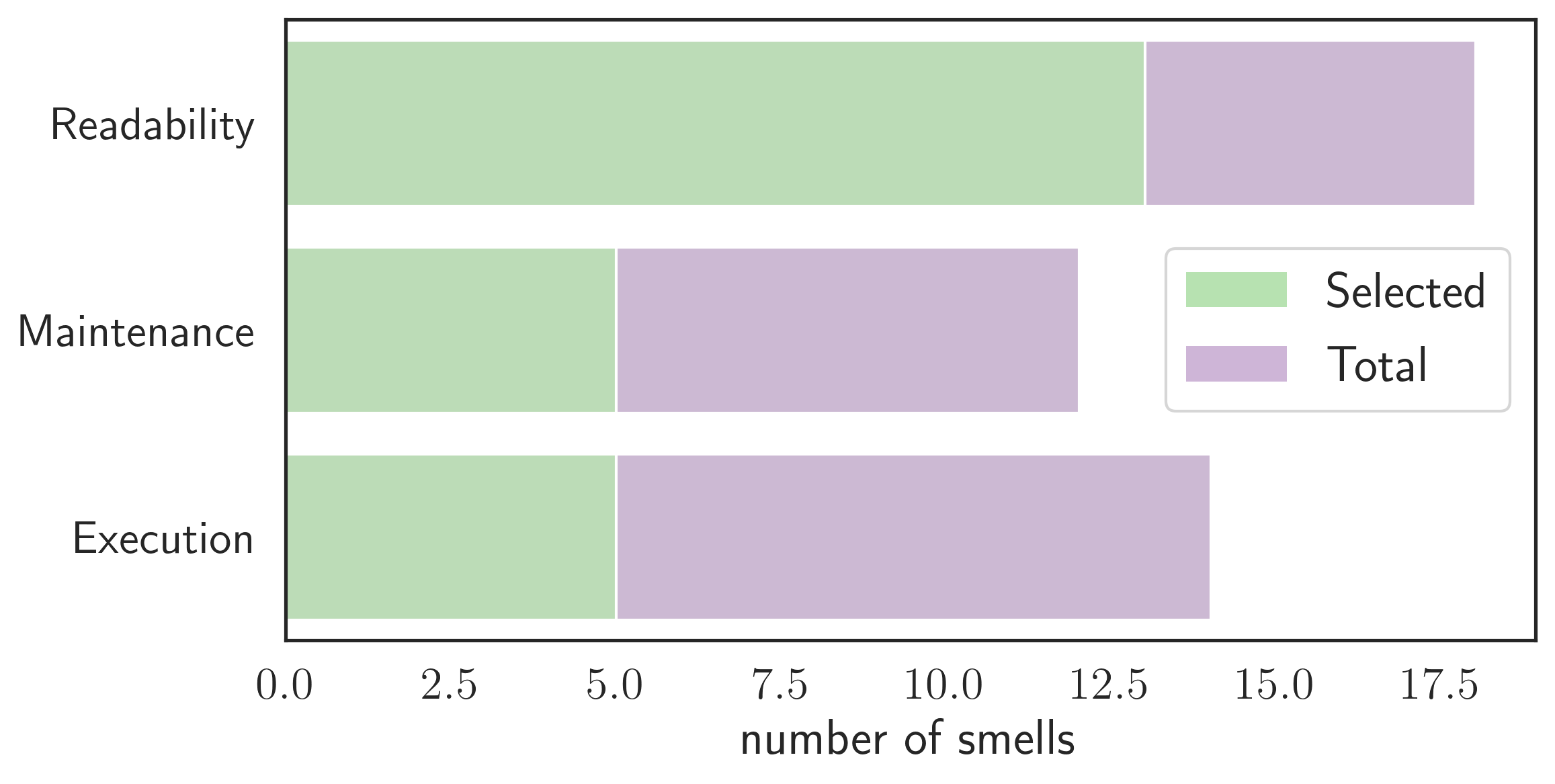}\label{fig:smell-issues}}
\caption{Properties of the identified SUIT smells. Figure~\ref{fig:smell-sources} displays the number of smells found in the academic and in the grey literatures. Figure~\ref{fig:smell-issues} displays the number of smells exhibiting a specific issue. Note that a smell can lead to several issues, hence the total is greater than the total number of SUIT smells.}  
\label{fig:smells}
\end{figure}

Figure~\ref{fig:smell-sources} shows the number of test smells that were found in academic and grey literature. We show both the 35 initial SUIT smells extracted from the literature and the subset of 16 SUIT smells for which we could extract a metric in the test code.

A SUIT smell is considered as covered if at least one of the smells grouped in the generalization step (Section~\ref{sec:experience-design-smells-collection}) is covered by the literature. If at least one source is covering a smell in the academic literature, we consider it as being covered by academia. Thus, smells labeled as originating from the grey literature are never mentioned in academic literature. The figure shows that while there exists an interest from practitioners with 35 unique smells discussed, the number of smells discussed in peer reviewed literature remains rather limited with only 10 smells identified. 
We also see that smells identified from the literature could generate a metric in six out of ten cases. The ones for which no metric could be derived are \emph{Inconsistent Wording} \citep{Hauptmann2013}, \emph{Unsuitable Naming} \citep{Chen2012}, \emph{Inconsistent Hierarchy} \citep{Hauptmann2013} and \emph{Data Creep} \citep{Alegroth2016b}. While \cite{Hauptmann2013} propose metrics for the smells they introduce, during our evaluation we observed a high rate of false positive for \emph{Inconsistent Wording} and \emph{Inconsistent Hierarchy} thus we excluded them for the final list. \cite{Chen2012} and \cite{Alegroth2016b} on the other hand do not propose a metric to automatically measure the smells they present. 

To conclude this analysis of Figure~\ref{fig:smell-sources}, we see that there exists a gap between the grey literature and the academic literature. We believe that more work need to be conducted to better understand and automatically detect and refactor SUIT smells in the academia. The present work is an attempt to fill this gap with the introduction of 15 new smells not previously studied in academia.

Figure~\ref{fig:smell-issues} presents the effects associated to each SUIT smell. Note that a smell can lead to issues from different categories. Thus, the total number of issues is greater than the number of smells. From the figure, we observe that although readability issues appear the most often with 18 instances, maintenance issues with 12 instances and execution issues with 14 instances do not fall far behind. Furthermore, to our surprise, SUIT smells affecting readability where the ones for which the most metrics could be computed. One reason explaining this phenomenon is that in the case of execution issues, information about the SUT and its execution are required. The same observation can be made for maintenance where the danger is coming from a divergence of the test from the SUT as it is the case for instance with the \emph{Lifeless} smell \citep{Buwalda2015, Renaudin2016, Buwalda2019} where the test is not following the same lifecycle as the application. Thus, in both cases, the extraction of the symptoms requires information about the structure and lifecycle of the SUT.

\begin{tcolorbox}[colframe=black!50!white, arc=10pt,before skip=10pt plus 2pt,after skip=10pt plus 2pt]
There is a gap between the academic and grey literatures in terms of SUIT smells, with 10 and 35 covered smells respectively. SUIT smells are reported to affect readability, maintenance, and test execution.
\end{tcolorbox}

\subsection{SUIT Smells Catalog}
\label{sec:results-smells-catalog}

The process described in Section~\ref{sec:experience-design-smells-collection} leads to a final list of 35 SUIT smells presented in Table~\ref{tab:smell-catalog}. 
Out of this list, we find that 16 smells can be automatically detected using metrics of the Robot Framework. 
Hence, we focus on these smells in our following questions.

\begin{table}
\centering

\begin{tabularx}{\textwidth}{>{\raggedright}p{1.5in}>{\raggedright}p{3in}}

\hline
\textbf{Name} & \textbf{Sources}\tabularnewline
\hline

\scriptsize{\textbf{Army of Clones}} & \scriptsize{\citep{Chen2012, Hauptmann2013, Hauptmann2015, Jones2019}} \tabularnewline \noalign{\vspace{4pt}}

\scriptsize{Complicated Setup Scenarios} & \citep{Scott2015} \tabularnewline \noalign{\vspace{4pt}}

\scriptsize{\textbf{Conditional Assertions}} & \scriptsize{\citep{Gawinecki2016}} \tabularnewline \noalign{\vspace{4pt}}

\scriptsize{Conspiracy Of Silence} & \scriptsize{\citep{Gawinecki2016, Sheth2020}} \tabularnewline \noalign{\vspace{4pt}}

\scriptsize{Data Creep} & \scriptsize{\citep{Alegroth2016b, Siminiuc2019, Shay2019}} \tabularnewline \noalign{\vspace{4pt}}

\scriptsize{Dependencies between tests} & \scriptsize{\citep{Klarck2014, Advolodkin2018, Cripsin2018, Bushnev2019, Goldberg2019}} \tabularnewline \noalign{\vspace{4pt}}

\scriptsize{Directly Executing UI Scripts} & \scriptsize{\citep{Scott2015}} \tabularnewline \noalign{\vspace{4pt}}

\scriptsize{Duplicate Check} & \scriptsize{\citep{Buwalda2019}} \tabularnewline \noalign{\vspace{4pt}}

\scriptsize{Eager Test} &  \scriptsize{\citep{England2016, Renaudin2016, Cripsin2018, Sciamanna2019, Temov2020}} \tabularnewline \noalign{\vspace{4pt}}

\scriptsize{\textbf{Hardcoded Environment}} & \scriptsize{\citep{Gawinecki2016, Sheth2020}} \tabularnewline \noalign{\vspace{4pt}}

\scriptsize{\textbf{Hiding Test Data}} & \scriptsize{\citep{Jain2007}} \tabularnewline \noalign{\vspace{4pt}}

\scriptsize{Implementation Dependent}  & \scriptsize{\citep{Jain2007, Kapelonis2018, Goldberg2019}} \tabularnewline \noalign{\vspace{4pt}}

\scriptsize{Inconsistent Hierarchy} & \scriptsize{\citep{Clayton2014, Gawinecki2016, Buwalda2019}} \tabularnewline \noalign{\vspace{4pt}}

\scriptsize{Inconsistent Wording} & \scriptsize{\citep{Hauptmann2013}} \tabularnewline \noalign{\vspace{4pt}}

\scriptsize{\textbf{Lack of Encapsulation}} & \scriptsize{\citep{Chen2012, Evangelisti2012, Klarck2014, Buwalda2015, England2016, Renaudin2016, Knight2017a, Goldberg2019, Jones2019, Shay2019}} \tabularnewline \noalign{\vspace{4pt}}

\scriptsize{Lack Of Early Feedback} & \citep{Dharmender2017} \tabularnewline \noalign{\vspace{4pt}}

\scriptsize{Lifeless} & \citep{Buwalda2015, Renaudin2016, Buwalda2019} \tabularnewline \noalign{\vspace{4pt}}

\scriptsize{\textbf{Long Test Steps}} & \scriptsize{\citep{Chen2012, Hauptmann2013, Buwalda2019}} \tabularnewline \noalign{\vspace{4pt}}

\scriptsize{\textbf{Middle Man}} & \scriptsize{\citep{Chen2012}} \tabularnewline \noalign{\vspace{4pt}}

\scriptsize{\textbf{Missing Assertion}} & \scriptsize{\citep{Klarck2014}} \tabularnewline \noalign{\vspace{4pt}}

\scriptsize{\textbf{Narcissistic}} & \scriptsize{\citep{England2016, Knight2017b}.} \tabularnewline \noalign{\vspace{4pt}}

\scriptsize{\textbf{Noisy Logging}} & \scriptsize{\citep{Jain2007}.} \tabularnewline \noalign{\vspace{4pt}}

\scriptsize{\textbf{Obscure Test}} & \scriptsize{\citep{Hauptmann2013, Gawinecki2016, Siminiuc2019}.} \tabularnewline \noalign{\vspace{4pt}}

\scriptsize{\textbf{On the Fly}} & \scriptsize{\citep{Archer2010}} \tabularnewline \noalign{\vspace{4pt}}

\scriptsize{\textbf{Over-Checking}} & \scriptsize{\citep{Buwalda2015, Renaudin2016}} \tabularnewline \noalign{\vspace{4pt}}

\scriptsize{Pointless Scenario Descriptions} & \scriptsize{\citep{England2016}} \tabularnewline \noalign{\vspace{4pt}}

\scriptsize{\textbf{Sensitive Locators}} & \scriptsize{\citep{Scott2015, Jones2019, Battat2020, Sheth2020}} \tabularnewline \noalign{\vspace{4pt}}

\scriptsize{\textbf{Sneaky Checking}} & \scriptsize{\citep{Kirkbride2014, Buwalda2015, Renaudin2016}} \tabularnewline \noalign{\vspace{4pt}}

\scriptsize{\textbf{Stinky Synchronization}} & \scriptsize{\citep{Gawinecki2016, Renaudin2016, Bushnev2019, Jones2019, Sheth2020}} \tabularnewline \noalign{\vspace{4pt}}

\scriptsize{Test Data Loss} & \scriptsize{\citep{Siminiuc2019}} \tabularnewline \noalign{\vspace{4pt}}

\scriptsize{Testing Data Not Code} & \scriptsize{\citep{Dharmender2017}} \tabularnewline \noalign{\vspace{4pt}}

\scriptsize{Unnecessary Navigation} & \scriptsize{\citep{Archer2010}} \tabularnewline \noalign{\vspace{4pt}}

\scriptsize{Unrealistic Data} & \scriptsize{\citep{Goldberg2019}} \tabularnewline \noalign{\vspace{4pt}}

\scriptsize{Unsecured Test Data} & \scriptsize{\citep{Morlion2019}} \tabularnewline \noalign{\vspace{4pt}}

\scriptsize{Unsuitable Naming} & \scriptsize{\citep{Chen2012, Goldberg2019, Shay2019, Sheth2020}} \tabularnewline \noalign{\vspace{4pt}}

\hline
\end{tabularx}

\caption{Catalog of SUIT Smells and their origin. Smell in bold are associated with a metric and a complete description is presented in Section~\ref{sec:results-smells-catalog}}
\label{tab:smell-catalog}.
\end{table}

Building on the tree model introduced in Section~\ref{sec:background-robot-framework} we represent each test $t$ as a rooted, ordered, directed, acyclic graph. The nodes, $N_t$, of the tree represent the set of calls to \emph{Keywords}, $C_t$ and the set of of arguments $A$ passed to the calls. A call, $c_t$ can be associated to a type namely: USER, INTERACTION, ASSERTION, CONTROLFLOW, GETTER, LOGGING and SYNC. Thus, the set of calls in a test performing an assertion is noted $C_{t, assertion}$. Finally, the definition of the \emph{User Kyeword} called by a test $t$ is noted $K_t$. Following this notation, we formally describe each smell for which a metric can be extracted with respect to its symptoms, its impact, and the metrics chosen to measure the prevalence of the symptoms following the protocol presented in Section~\ref{sec:smell-detection}. Finally, we present the refactoring actions removing the symptoms from the test codebase.

\subsubsection{Army of Clones (AoC)}

\paragraph{Description:}

Different tests perform and implement similar actions, leading to duplicated pieces of test code.

\paragraph{Impact on Readability:} 

Test sequences which are similar but not identical are not easy to distinguish. It is not easy to grasp the intention of the test in comparison with its clone.

\paragraph{Impact on Maintenance:} 

The effort to maintain duplicated parts of tests increases. Furthermore, it is difficult to determine where maintenance has to be performed.

\paragraph{Detection Method:}

Code duplication can be observed at different levels. Here, for the body of a \emph{User Keyword}, we detect if there exists a clone of type 1 (code duplication at the token level) or type 2 (code duplication at syntax level allowing for minor syntactic changes such as variables name) in the test suite. Thus, we express the count metric, $S_{AoC}(t)$, as the number of calls to \emph{User Keywords} that have a clone and the density metric, $D_{AoC}(t)$, as the number of calls to \emph{User Keywords} that have a clone over the total number of unique \emph{User Keywords} called by the test. More formally:

\begin{equation*}
    S_{AoC}(t) = \abs{K_t \cap K_{clone}} 
\end{equation*}

\begin{equation*}
    D_{AoC}(t) = \frac{\abs{K_t \cap K_{clone}}}{\abs{K_{t}}}
\end{equation*}

where $K_{t}$ is the set of unique \emph{User Keywords} called by test $t$ and $K_{clone}$ is the set of unique \emph{User Keywords} that have at least one clone in the test suite.

\paragraph{Refactoring Actions:}

The symptom is considered refactored if the \emph{User Keyword} that is called by a test  and have at least one clone in the test suite, $k_{t, clone}$, is removed. Thus, we propose the refactoring pattern, $R_{AoC}(k)$, as follow:

\begin{equation*}
    R_{AoC}(k) = k_{t, clone} \xrightarrow{action} \emptyset
\end{equation*}

\subsubsection{Conditional Assertions (CA)}

\paragraph{Description:}

The test verifies different properties depending on the environment when the environment state may change from one execution to the next.
\paragraph{Impact on Readability:} 

With more complex logic in the assertions, it becomes harder to capture their meaning.

\paragraph{Impact on Execution:} 

More complex code might introduce bugs in the test code.

\paragraph{Detection Method:}

We consider assertions nodes, $C_{assertion}$, to be symptomatic if they have a parent node which is a conditional node and have no sibling nodes in the call graph, $C_{condition}$. Thus, we express the count metric, $S_{CA}(t)$, as the number of conditional assertion calls and the density metric, $D_{CA}(t)$, as the number of conditional assertion calls over the total number of assertion calls.

\begin{equation*}
    S_{CA}(t) = \abs{C_{t} \cap C_{assertion} \cap C_{condition}}
\end{equation*}

\begin{equation*}
    D_{CA}(t) = \frac{\abs{C_{t} \cap C_{assertion} \cap C_{condition}}}{\abs{C_{t} \cap C_{assertion}}}
\end{equation*}

where $\abs{C_{t} \cap C_{assertion}}$ is the size of the set of calls to \emph{Library Keyword} annotated as ``assertion'' for a test $t$ and $\abs{C_{t} \cap C_{assertion} \cap C_{condition}}$  is the size of the set of calls to \emph{Library Keyword} annotated as ``assertion'' for which the caller is a conditional node that has only one child (logging nodes excluded).

\paragraph{Refactoring Action:}

The symptom is considered refactored if the conditional assertion node is removed from the call graph. Thus, we accept the following refactoring pattern, $R_{CA}(c)$, as removing a symptom in a node $c$:

\begin{equation*}
    R_{CA}(c) =  c_{t, assertion, condition} \xrightarrow{action} c_{t, assertion}
\end{equation*}

The assertion $c_{t, assertion}$ replaces its former parent node $c_{t, assertion, condition}$. Note that removing a parent of $c_{t, assertion, condition}$ or adding a sibling to the child assertion node are not considered as fixing the symptom.












\subsubsection{Hardcoded Environment (HE)}

\paragraph{Description:}

The test contains hardcoded references to the environment when the same requirement must be run against different test environments instead of having an environment-agnostic test.

\paragraph{Impact on Maintenance:} 

Updating the configuration requires modifying multiple locations in different tests.

\paragraph{Detection Method:}

The metric we propose covers the case of multi-browser configuration. Here, when a browser is loaded, the metric ensure that the web-driver is not instantiated with a hardcoded configuration for the browser. Thus, we express the count metric, $S_{HE}(t$, as the number of configuration arguments that are hardcoded and the density metric, $N_{HE}(t)$, as he number of configuration arguments that are hardcoded over the total number of configuration arguments. More formally:

\begin{equation*}
    S_{HE}(t) = \abs{A_{t} \cap A_{config} \cap A_{hardcoded}}
\end{equation*}

\begin{equation*}
    D_{HE}(t) = \frac{\abs{A_{t} \cap A_{config} \cap A_{hardcoded}}}{\abs{A_{t} \cap A_{config}}}
\end{equation*}

where $\abs{A_{t} \cap A_{config}}$ is the size of the set of arguments in calls to \emph{Library Keywords} annotated as ``configuration'' in a test $t$ and $\abs{A_{t} \cap A_{config} \cap A_{hardcoded}}$ is the size of the set of arguments in calls to \emph{Library Keywords} annotated as ``configuration'' for which the value is hardcoded.

\paragraph{Refactoring Action:}

The symptom is considered refactored if the hardcoded argument in a call to a \emph{Library Keyword} annotated as ``configuration'', $a_{t, config, hardcoded}$, is replaced with a variable, $a_{t, config, variable}$. Thus, we propose the following refactoring pattern, $R_{HE}(a)$, for an argument $a$:

\begin{equation*}
    R_{HE}(a) = a_{t, config, hardcoded} \xrightarrow{action}  a_{t, config, variable}
\end{equation*}

\subsubsection{Hidden Test Data (HTD)}

\paragraph{Description:}

The data are not directly visible and understandable in the test but are hidden in the fixture code.

\paragraph{Impact on Readability:} 

The data is completely obscure to the future reader making the intent of the test difficult to understand.

\paragraph{Detection Method:}

In this work, we associate the fixture code to the setup of a test. We consider data access as reading input from external resources through a \emph{Library Keyword} annotated as ``getter''. Thus, we express the count metric, $S_{HTD}(t)$, as the number of calls to getter in the setup of a test and $D_{HTD}(t)$ as the number of calls to getter in the setup of a test over the total number of calls in the setup of that test. More formally:

\begin{equation*}
    S_{HTD}(t) = \abs{C_{t} \cap C_{setup} \cap C_{getter}}
\end{equation*}

\begin{equation*}
    D_{HTD}(t) = \frac{\abs{C_{t} \cap C_{setup} \cap C_{getter}}}{\abs{C_{t} \cap C_{setup}}}
\end{equation*}

where $\abs{C_{t} \cap C_{setup}}$ is the size of the set of calls to \emph{Library Keywords} in the setup of test $t$ and $\abs{C_{t} \cap C_{setup} \cap C_{getter}}$ is the size of the set of calls to \emph{Library Keywords} annotated as ``getter'' in the setup of test $t$.

\paragraph{Refactoring Action:}

The symptom is considered refactored if the call to the \emph{Library Keywords} annotated as ``getter'' in the setup of test $t$, $c_{t, setup, getter}$, is removed. Thus, we propose the following refactoring pattern, $R_{HTD}(c)$ performed on a \emph{Library Keyword} call $c$ as follow:

\begin{equation*}
    R_{HTD}(c) = c_{t, setup, getter} \xrightarrow{action}  \emptyset
\end{equation*}

Note that removing a parent node of $c_{t, setup, getter}$ is not considered as a fix.

\subsubsection{Lack of Encapsulation (LoE)}

\paragraph{Description:}

The implementation details of a test are not properly hidden in the implementation layer and start appearing in its acceptance criteria.

\paragraph{Impact on Readability:} 

The acceptance criteria is meant to convey intention over implementation. Focusing on implementation in the acceptance criteria makes the intent harder to grasp.

\paragraph{Detection Method:}

Typically the acceptance criteria makes call to the implementation layer which subsequently relies on the application driver layer. The metric detects the direct calls from the acceptance criteria (test steps) to \emph{Library Keywords}. Thus, we express the count metric, $S_{LoE}(t)$, as the number of direct calls to a driver from the acceptance criteria of a test and the density metric, $D_{LoE}(t)$, as the number the number of direct calls to a driver from the acceptance criteria of a test over the total number of steps of the acceptance criteria of a test. More formally:

\begin{equation*}
    S_{LoE}(t) = \abs{C_{t} \cap C_{step} \cap C_{driver}}
\end{equation*}

\begin{equation*}
    D_{LoE}(t) = \frac{\abs{C_{t} \cap C_{step} \cap C_{driver}}}{\abs{C_{t} \cap C_{step}}}
\end{equation*}

where $\abs{C_{t} \cap C_{step}}$ is the size of the set of steps in the acceptance criteria of the test $t$ and $\abs{C_{t} \cap C_{step} \cap C_{driver}}$ is the size of the set of steps in the acceptance criteria of the test $t$ directly calling the application driver, \emph{i.e.} a (\emph{Library Keyword}).

\paragraph{Refactoring Action:}

The symptom is considered refactored if the direct call to a \emph{Library Keyword} is removed from the acceptance criteria. Thus, we propose the refactoring patterns, $R_{LoE, 1}(c)$ and $R_{LoE, 2}(c)$, performed on a \emph{Library Keyword} call $c$ as follow:

\begin{equation*}
    R_{LoE, 1}(c) = c_{t, step, driver} \xrightarrow{action_1} c_{t, step, \neg driver}
\end{equation*}

\begin{equation*}
    R_{LoE, 2}(c) = c_{t, step, driver} \xrightarrow{action_2} \emptyset
\end{equation*}

In the first equation, $R_{LoE, 1}(c)$, the direct call to a \emph{Library Keyword} in the acceptance criteria is replaced by a call to a \emph{User Keyword} where as in the second equation, $R_{LoE, 2}(c)$, the call is removed.

\subsubsection{Long Test Steps (LTS)}

\paragraph{Description:}

One or many test steps are very long, performing a lot of actions on the system under test.

\paragraph{Impact on Readability:} The intention of the step is difficult to grasp.

\paragraph{Impact on Execution:} With each action on the system under test, there is a chance of something going wrong. The higher this number, the more fragile the test becomes.

\paragraph{Detection Method:}

For a test step, $K_{t, step}$, of a test $t$, the metric counts the number of \emph{Library Keyword} annotated as ``action'' (triggering an event on the SUT) called directly or indirectly by $K_{t, step}$. If the value is greater than a threshold $L$, then $K_{t, step}$ is considered symptomatic. Thus, we express the count metric, $S_{LTS}(t)$, as the number of steps that are performing a number of actions greater than a threshold and the density metric, $D_{LTS}(t)$, s the number of steps that are performing a number of actions greater than a threshold over the total number of steps. More formally:

\begin{equation*}
    S_{LTS}(t) = \abs{C_t \cap C_{step \geq L}}
\end{equation*}

\begin{equation*}
    D_{LTS}(t) = \frac{\abs{C_t \cap C_{step \geq L}}}{\abs{C_t \cap C_{step}}}
\end{equation*}

where $\abs{C_t \cap C_{step}}$ is the number of steps in a test $t$ and $\abs{C_t \cap C_{step \geq L}}$ is the number of steps calling more than $L$ \emph{Library Keyword} annotated as ``action''. 

\begin{figure}
\centering
\includegraphics[width=0.6\linewidth]{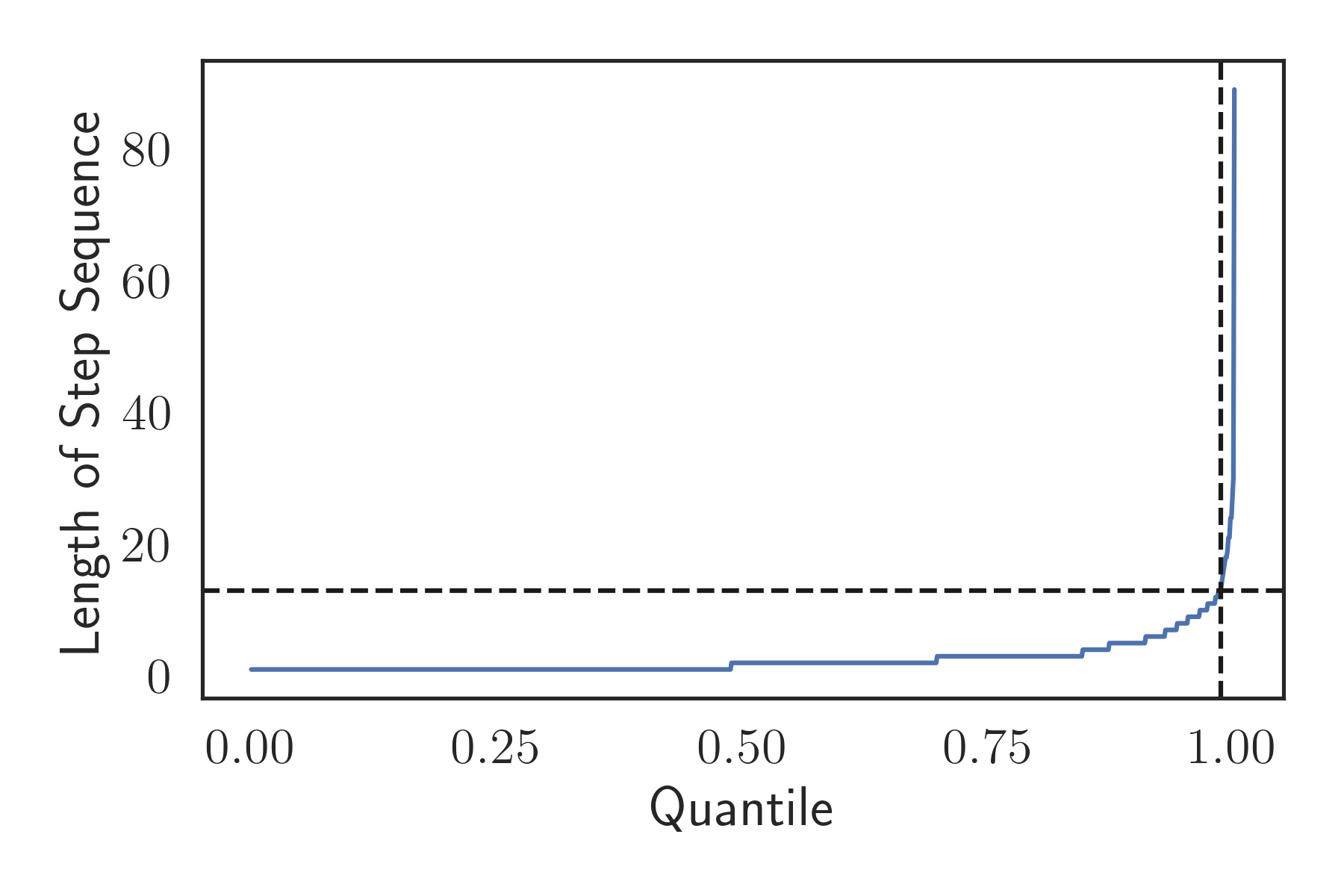}
\caption{The blue curve represent the evolution of the length of the sequences of the step as the quantiles increase. The intersection of the black doted lines displays the knee point (0.986, 13) at which the values sequence lengths values start to increase dramatically.}  
\label{fig:step-sequences-quantiles}
\end{figure}

Because the parameter $L$ needs to be set empirically, we compute a deviation threshold based on the distribution of our dataset. Using the analysis of the evolution of the quantiles, we compute at which point the values start to rapidly deviates by computing the knee curve of the quantiles distribution function proposed by \cite{Satopaa2011}. Following this approach we find the knee point at 13 actions on the SUT for a step (quantile = 0.986, see Figure~\ref{fig:step-sequences-quantiles}). Therefore, we set $L = 13$  and consider any step performing a sequence of actions on the SUT greater than 13 to be too long.

\paragraph{Refactoring Action:}

The symptom is considered refactored if the number of actions performed on the SUT by a long step sees its value pass under the threshold $L$. We do not specify how the calls on the SUT have to be transformed, as long as the test step call is left unchanged. Thus, we propose the refactoring pattern, $R_{LTS}(c)$, where $c$ is a long test step:

\begin{equation*}
    R_{LTS}(c) = c_{t, step\geq L} \xrightarrow{action} c_{t, step < L}
\end{equation*}

where $c_{t, step\geq L}$ is a step yielding at least $L$ actions on the SUT while $c_{t, step < L}$ is a step yielding less than $L$ actions on the SUT.

\subsubsection{Middle Man (MM)}

\paragraph{Description:}

A test component (keyword, macro, function) delegates all its tasks to another test component.

\paragraph{Impact on Readability:} The levels of indirection make the test harder to follow by future readers.

\paragraph{Detection Method:}

The principle of delegating is doing nothing and simply calling another function. Thus, we express the count metric, $S_{MM}(t)$, as the number of \emph{User Keywords} called from a test that are composed of a single call to another \emph{Keyword} and the density metric, $D_{MM}(t)$, as the he number of \emph{User Keywords} called from a test that are composed of a single call to another \emph{User Keyword} and the density metric over the total number of \emph{Keyword} called by the test. More formally:

\begin{equation*}
    S_{MM}(t) = \abs{K_{t} \cap K_{delegate}}
\end{equation*}

\begin{equation*}
    D_{MM}(t) = \frac{\abs{K_{t} \cap K_{delegate}}}{\abs{K_{t}}}
\end{equation*}

where $\abs{K_{t}}$ is the number of \emph{User Keywords} in a test $t$ and $\abs{K_{t} \cap K_{delegate}}$ is the number of \emph{User Keywords} performing a single call to one other \emph{User Keyword} (we ignore logging action) without performing any subsequent action on their own, \emph{i.e.} \emph{Delegate Keyword}.

\paragraph{Refactoring Actions:}

The symptom is considered refactored if the \emph{Delegate Keyword} call is replaced with another call which is not simply delegating its actions. Thus, we propose the refactoring pattern, $R_{MM}(c)$ where $c$ is a call to a \emph{Delegate Keyword} as:

\begin{equation*}
    R_{MM}(c) = c_{K_{t, delegate}} \xrightarrow{action} c_{K_{t, \neg delegate}}
\end{equation*}

where $c_{K_{t, delegate}}$ is a call to a \emph{Delegate Keyword} and $c_{K_{t, \neg delegate}}$ is a call to any \emph{Keyword} not performing delegation.

\subsubsection{Missing Assertion (MA)}

\paragraph{Description:}

The test lacks any explicit assertions.

\paragraph{Impact on Readability:} 

Future readers are left in the potentially frustrating position of puzzling over the intention of the test.

\paragraph{Detection Method:}

The metric detects the absence of call to \emph{Library Keyword} annotated as ``assertion'' within the test. Because the symptom can at most appear once in the test, both the count metric, $S_{MA}(t)$ and the density metric $D_{MA}(t)$ are express as the absence of an assertion call in the test. More formally:

\begin{equation*}
    S_{MA}(t) = D_{MA}(t) = 
    \begin{cases}
        1,   & \text{if } C_{t} \cap C_{assert} = \emptyset\\
        0,   & \text{otherwise}
    \end{cases}
\end{equation*}

\paragraph{Refactoring Actions:}

The symptom is considered refactored if \emph{Library Keyword} annotated as ``assertion'' is introduced in test that is missing any assertion. Thus, we propose the refactoring pattern, $R_{MA}(t)$, where $t$ is the test lacking any assertion as:

\begin{equation*}
    R_{MA}(t) = \emptyset \xrightarrow{action} c_{t, assertion}
\end{equation*}

where $c_{t, assertion}$ is a call to a \emph{Library Keyword} annotated as ``assertion'' in a test $t$.

\subsubsection{Narcissistic (N)}

\paragraph{Description:}

The test uses the first person pronoun ``I'' to refer to its actors and does not uniquely qualify those actors.

\paragraph{Impact on Readability:} The test is harder to read because it is not clear who ``I'' is and what are the roles that ``I'' has.

\paragraph{Detection Method:}

Using text tagging, we identify calls from the acceptance criteria, \emph{i.e.} ``test steps'', using a personal pronouns as the subject in there name as symptomatic. Furthermore, the implementation used in our experiments supports the different languages present in our dataset (namely, French and English).. Thus, we express the count metric, $S_{N}(t)$, as the number of steps in the acceptance criteria of a test that are using a personal pronoun and the density metric, $D_{N}(t)$, as the number of steps in the acceptance criteria of a test that are using a personal pronoun over the total number of steps in the acceptance criteria of the test:

\begin{equation*}
    S_{N}(t) = \abs{C_{t} \cap C_{step} \cap C_{I}}
\end{equation*}

\begin{equation*}
    D_{N}(t) = \frac{\abs{C_{t} \cap C_{step} \cap C_{I}}}{\abs{C_{t} \cap C_{step}}}
\end{equation*}

where  $\abs{C_{t} \cap C_{step}}$ is the number of ``test steps'' for a test $t$ and $\abs{C_{t} \cap C_{step} \cap C_{I}}$ is the number of ``test steps'' using a personal pronouns as the subject in there name.

\paragraph{Refactoring Actions:}

The symptom is considered refactored if the name of a symptomatic ``test steps'' is changed so that it does not contain a personal pronoun anymore. Thus,  we propose the refactoring pattern, $R_{N}(c)$, where $c$ is a step of the acceptance criteria of a test $t$:

\begin{equation*}
    R_{N}(c) = c_{t, step, I} \xrightarrow{action} c_{t, step, \neg I}
\end{equation*}

where $c_{t, step, I}$ is a \emph{User Keyword} called by a ``test steps'' using the personal pronoun ``I'' as the subject and $c_{t, step, \neg I}$ is the \emph{User Keyword} with its new name not using the personal pronoun ``I'' as the subject. Therefore, a fix is detected only when the name of a \emph{User Keyword} called by a ``test steps'' is changed.

\subsubsection{Noisy Logging (NL)}

\paragraph{Description:}

The test logs the state of the fixtures.

\paragraph{Impact on Execution:} 

There is too much noise in the output from the tests, making its analysis more cumbersome.

\paragraph{Detection Method:}

In this work, we associate the fixture code to the setup of a test. Thus, we express the count metric, $S_{NL}(t)$, as the number of calls to \emph{Library Keywords} annotated as ``logging'' from the setup of a test and the density metric, $D_{NL}(t)$, as the number of calls to \emph{Library Keywords} annotated as ``logging'' from the setup of a test over the total number of calls from the setup of the test. More formally:

\begin{equation*}
    S_{NL}(t) = \abs{C_{t} \cap C_{setup} \cap C_{logging}}
\end{equation*}

\begin{equation*}
    D_{NL}(t) = \frac{\abs{C_{t} \cap C_{setup} \cap C_{logging}}}{C_{t} \cap C_{setup}}
\end{equation*}

where $\abs{C_{t} \cap C_{setup}}$ is the size of the set of \emph{Library Keyword} called from the setup of a test $t$ and $\abs{C_{t} \cap C_{setup} \cap C_{logging}}$  is the size of the set of \emph{Library Keyword} annotated as ``logging'' called from the setup of test $t$.

\paragraph{Refactoring Action:}

The symptom is considered refactored if the call to the \emph{Library Keyword} annotated as ``logging'' called from the setup of test $t$, $c_{t, setup, log}$, is removed. Thus we propose the refactoring pattern, $R_{NL}(c)$, performed on a \emph{Keyword} call $c$ as follow:

\begin{equation*}
    R_{NL}(c) = c_{t, setup, log} \xrightarrow{action} \emptyset
\end{equation*}

Note that removing a parent of $c_{t, setup, log}$ is not considered as fixing the smell.

\subsubsection{Obscure Test (OT)}

\paragraph{Description:}

The test behavior is difficult to understand because the test does not clearly state what it is verifying. Typical symptoms are hardcoded values, high cyclomatic complexity and/or function or procedure calls with high number of parameters.

\paragraph{Impact on Readability:} Future reader might not understand the meaning of a hardcoded value, hence, missing the intention of the test. As with a high cyclomatic complexity it becomes hard for the future reader to follow the execution flow of the test and grasp what it is doing.

\paragraph{Impact on Maintenance:} It is difficult to determine where to perform changes if hardcoded values are scattered all over the test code. Furthermore, test with high cyclomatic complexity might have side effect overseen during maintenance which might lead to future problems.

\paragraph{Detection Method:}

In this work, we focus on one of the expression of an obscure test: the overuse of hardcoded values. The code starts to smell when hardcoded values are used directly in calls to both \emph{User Keyword} and \emph{Library Keywords}, instead of relying on variables. Thus, we express the count metric, $S_{OT}(t)$, as the number of hardcoded arguments present in a test and the density metric $D_{OT}(t)$ as the number of hardcoded arguments present in the test over the total number of arguments from that test. More formally:

\begin{equation*}
    S_{OT}(t) = \abs{A_{t} \cap A_{hardcoded}}
\end{equation*}

\begin{equation*}
    D_{OT}(t) = \frac{\abs{A_{t} \cap A_{hardcoded}}}{\abs{A_t}}
\end{equation*}

where $\abs{A_t}$ is the size of the set of arguments passed to \emph{Keyword} calls in a test $t$ and $\abs{A_{t} \cap A_{hardcoded}}$ is the size of the set of arguments passed to \emph{Keyword} calls which are hardcoded.

\paragraph{Refactoring Actions:}

Focusing on hardcoded values, the symptom is considered refactored if an argument passed to \emph{Keyword} call as a hardcoded value is replaced by a variable. Thus, we propose the refactoring pattern, $R_{OT}(a)$, where $a$ is a \emph{Keyword} call arguments as:

\begin{equation*}
    R_{OT}(v) = a_{t, hardcoded} \xrightarrow{action} a_{t, variable}
\end{equation*}

where $a_{t, hardcoded}$ is a hardcoded \emph{Keyword} call argument and $a_{t, variable}$ is a variable \emph{Keyword} call argument.

\subsubsection{On the Fly (OtF)}
\paragraph{Description:}

The test calculates an expected results during its execution instead of relying on pre-computed values.

\paragraph{Impact on Readability:} 

By embedding the business rule in the assertion, the code for the automated test can become as complicated as the system under test.

\paragraph{Detection Method:}

The expected value should be a constant or a reference to a constant and not computed during the test. Thus, we express the count metric, $S_{OtF}(t)$, as the number of expected values that are computed in the test and the density metric, $D_{OtF}(t)$, as the number of expected arguments from assertion calls that are computed in the test over the total number of expected arguments from assertion calls:

\begin{equation*}
    S_{OtF}(t) = \abs{A_{t} \cap A_{expected} \cap A_{computed}}
\end{equation*}

\begin{equation*}
    D_{OtF}(t) = \frac{\abs{A_{t} \cap A_{expected} \cap A_{computed}}}{\abs{A_{t} \cap A_{expected}}}
\end{equation*}

where $\abs{A_{t} \cap A_{expected}}$ is the number of ``expected'' arguments in calls to \emph{Library Keywords} annotated as ``assertion'' for a test $t$ and  $\abs{A_{t} \cap A_{expected} \cap A_{computed}}$ is the number of ``expected'' arguments in calls to \emph{Library Keywords} annotated as ``assertion'' resolved during the execution of the test $t$. Note that the identification of the ``expected'' argument is based on the definition of the \emph{Library Keyword}. When a \emph{Library Keyword} annotated as ``assertion'' contains a field called ``expected'', it is considered by the engine as the placeholder for the expected value. For instance, the library keyword \emph{Should be equal} from the Builtin library takes six arguments: \emph{value}, \emph{expected}, \emph{message}, \emph{values}, \emph{ignore\_case} and \emph{formatter}. Hence, in this case the expected argument is the second one.

\paragraph{Refactoring Action:}

The symptom is considered refactored when the assertion is preserved but the expected value is not computed on the fly. This leads to the following equation for a refactoring action, $R_{OtF}$, addressing a symptom in an argument $a$:

\begin{equation*}
    R_{OtF}(a) =  a_{t, expected, computed} \xrightarrow{action} a_{t, expected, \neg computed}
\end{equation*}

where $a_{t, expected, computed}$ is an expected value that is computed on the fly and $a_{t, expected, \neg computed}$ is an expected value that is not computed on the fly, being either hardcoded or provided through a variable pointing to a static value. Thus removing the assertion would not be considered as removing the symptom since $a_{t, expected, \neg computed}$ would not be present.

\subsubsection{Over-Checking (OC)}

\paragraph{Description:}

The test performs some assertions that are not relevant for its scope.

\paragraph{Impact on Readability:} 

It becomes harder to understand what is the main intent of the test. Many assertions suggests the test is checking many different properties.

\paragraph{Impact on Maintenance:} 

The test may be too sensitive to the evolution of the SUT, verifying implementation properties instead of behavioral ones.

\paragraph{Detection Method:}

As the ratio assertions to actions on SUT increases, the chances that all the assertions are relevant decreases. Thus we express the count metric, $S_{OC}(t)$, as the number of assertions in a test and the density metric, $N_{OC}(t)$, as the number of assertions in a test over the total number of calls in the test. More formally:

\begin{equation*}
    S_{OC}(t) = \abs{C_{t} \cap C_{assertion}}
\end{equation*}

\begin{equation*}
    N_{OC}(t) = \frac{ \abs{C_{t} \cap C_{assertion}}}{\abs{C_{t}}}
\end{equation*}

where $\abs{C_{t}}$ is the size of the set \emph{Library Keywords} calls and $\abs{C_{t} \cap C_{assertion}}$ is the size of the set of calls to \emph{Library Keywords} annotated as ``assertion''.

\paragraph{Refactoring Actions::}

The symptom is considered refactored if the call to a \emph{Library Keyword} annotated with ``assertion'', $c_{t, assertion}$ , is removed from a test $t$. Thus, we propose the refactoring pattern, $R_{OC}(c)$ where $c$ is a call to an \emph{Library Keyword} annotated with ``assertion'' as:

\begin{equation*}
    R_{OC}(c) = c_{t, assertion} \xrightarrow{action} \emptyset
\end{equation*}

\subsubsection{Sensitive Locators (SL)}

\paragraph{Description:}

The test uses element identification selectors that have long chains to identify an element in the user interface. e.g. complex x-pass or CSS selector for web application.

\paragraph{Impact on Maintenance:} This leads to fragile tests, as any change in that chain from the user interface representation will break the tests.

\paragraph{Detection Method:}

The complexity of a locator can be expressed by how deep the locator needs to go in the hierarchy of the UI, be it an x-pass, a CSS selector or any UI representation based on a hierarchy. A locator, $A_{locator}$, can be expressed as the number of GUI element, $\abs{E}$, that have to be traversed to uniquely locate the target GUI element. Thus, we express the count metric, $S_{SL}(t)$, as the number of locator arguments that visit more than one GUI elements and the density metric, $D_{SL}(t)$, the number of of locator arguments that visit more than one GUI elements in a test over the total number of locator arguments present in the test. More formally:

\begin{equation*}
    S_{SL}(t) = \abs{A_t \cap A_{locator} \cap A_{\abs{E} > 1}}
\end{equation*}

\begin{equation*}
    D_{SL}(t) = \frac{\abs{A_t \cap A_{locator} \cap A_{\abs{E} > 1}}}{\abs{A_t \cap A_{locator}}}
\end{equation*}

where $\abs{A_t \cap A_{locator} \cap A_{\abs{E} > 1}}$ is the number of locators that require to visit more than one element $E$ of the GUI to be uniquely identified (\emph{e.g.} the XPath ``/html/body/div[4]/button'' visits 4 elements to quality the button where the XPath ``//button[@id ="unique-id"]'' only needs to visit one element). Note that Robot Framework using dynamic types, only \emph{Library Keyword} calls explicitly specify a type for their parameters. Therefore, for each \emph{Library Keyword} call requiring a locator as an argument, the engine resolve all the values possible for the argument within the test to populate the set $A_{locator}$.

\paragraph{Refactoring Actions:}

The symptom is considered refactored if the value of a node $l$ flagged as complex locator sees its length go down to one. Thus, we propose the refactoring pattern, $R_{SL}(l)$, as follow:

\begin{equation*}
    R_{SL}(c) = l_{\abs{E} > 1} \xrightarrow{action} l_{\abs{E} = 1}
\end{equation*}

where $l_{\abs{E} > 1}$ is a node defining the value of a sensitive locator and $l_{\abs{E} = 1}$ is the same node but with a simple locator expression. Note that a change is only accounted for when the value of the locator is modified.

\subsubsection{Sneaky Checking (SC)}

\paragraph{Description:}

The test hides its assertions in actions that are at the wrong level of details.

\paragraph{Impact on Readability:} The future reader is not able to understand what is being tested by just looking at the main steps of the acceptance criteria without a need to inspect how low level details are implemented.

\paragraph{Detection Method:}

A \emph{User Keyword} only calling an \emph{Library Keyword} annotated as ``assertion'' can be seen as hiding the assertion to the callers of that \emph{User Keyword}. Thus, we express the count metric, $S_{SC}(t)$, as the number of unique \emph{User Keywords} called by a test and only calling an assertion and the density metric, $D_{SC}(t)$, as the number of unique \emph{User Keywords} called by the test and only calling an assertion over the number of unique \emph{User Keywords} called by that test. More formally:

\begin{equation*}
    S_{SC}(t) = \abs{K_{t} \cap K_{assert}}
\end{equation*}

\begin{equation*}
    D_{SC}(t) = \frac{\abs{K_{t} \cap K_{assert}}}{\abs{K_{t}}}
\end{equation*}

where $\abs{K_{t}}$ is the total number of unique \emph{User Keywords} and $\abs{K_{t} \cap K_{assert}}$ is the number of \emph{User Keywords} only calling an \emph{Library Keyword} annotated as ``assertion'' (logging actions are ignored).

\paragraph{Refactoring Actions:}

The symptom is considered refactored if a \emph{User Keywords} only calling a \emph{Library Keyword} annotated as ``assertion'', $k_{t, assert}$, is removed from the test $t$. Thus, we propose the refactoring pattern, $R_{SC}(k)$, where $k$ is a \emph{User Keywords} as:

\begin{equation*}
    R_{SC}(k) = k_{t, assert} \xrightarrow{action} \emptyset
\end{equation*}

\subsubsection{Stinky Synchronization (SS)}

\paragraph{Description:}

The test fails to use proper synchronization points with the system under test.

\paragraph{Impact on Execution:} 

The test becomes oversensitive to the response time, leading to flaky tests, or very slow tests when choosing very conservative wait points.

\paragraph{Detection Method:}

This symptom is associated with the use of explicit and fixed synchronization, independent from the SUT such as a pausing the test for a specific amount of time. Thus, we express the count metric, $S_{SS}(t)$, as the number of calls to explicit pause from a test and the density metric, $D_{SS}(t)$, as the number of calls to explicit pause from a test over the total number of synchronization calls. More formally:

\begin{equation*}
    S_{SS}(t) = \abs{C_{t} \cap C_{sync} \cap C_{sleep}}
\end{equation*}

\begin{equation*}
    D_{SS}(t) = \frac{\abs{C_{t} \cap C_{sync} \cap C_{sleep}}}{\abs{C_{t} \cap C_{sync}}}
\end{equation*}

where $\abs{C_{t} \cap C_{sync}}$ is the size of the set of calls to \emph{Library Keyword} annotated as ``synchronization'' in a test $t$ and $\abs{C_{t} \cap C_{sync} \cap C_{sleep}}$ is the size of the set of calls to \emph{Library Keyword} annotated as ``synchronization'' by pausing the test execution for a specified amount of time. In the case of Robot Framework, it is instantiated by calls to the \emph{Library Keyword} ``Sleep''.

\paragraph{Refactoring Actions:}

The symptom is considered refactored if a \emph{Library Keyword} call annotated as ``synchronization'' which pausing the test execution of the test for a specified amount of time, $c_{t,sync,sleep}$, is removed or replaced by another \emph{Library Keyword} call annotated as ``synchronization'', $c_{t,sync, \neg sleep}$. Thus, we propose two refactoring patterns, $R_{SS, 1}(c)$ and $R_{SS, 2}(c)$, as follow:

\begin{equation*}
    R_{SS, 1}(c) = c_{t,sync,sleep} \xrightarrow{action_1} \emptyset
\end{equation*}

\begin{equation*}
    R_{SS,2}(c) = c_{t,sync,sleep} \xrightarrow{action_2} c_{t,sync, \neg sleep}
\end{equation*}

\subsection{RQ2: SUIT Smells Distribution}
\label{sec:results-smells-diffusion}

\begin{figure}
\centering
\includegraphics[width=0.9\linewidth]{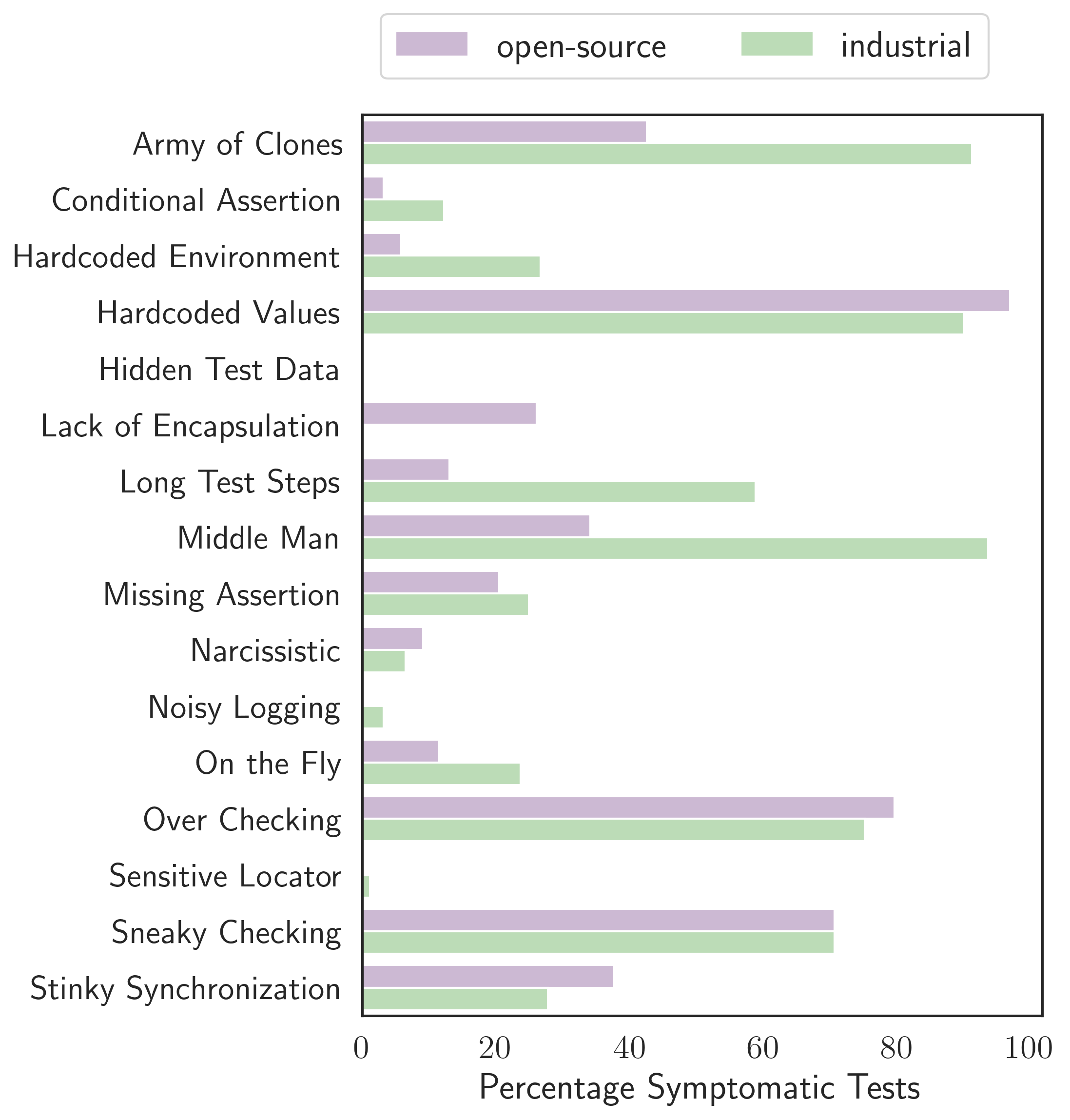}
\caption{Percentage of tests exhibiting at least one symptom in open-source and industrial projects.}  
\label{fig:diffusion}
\end{figure}
Building on the metrics derived in Section~\ref{sec:results-smells-catalog}, we show in Figure~\ref{fig:diffusion} the number of tests exhibiting at least one smell symptom. The smell \emph{Hidden Test Data} does not in any of the studied projects. Furthermore, the smells \emph{Noisy Logging}, \emph{Sensitive Locator}, \emph{Narcissistic} and \emph{HardCoded Environment} manifest in less than 10\% of the tests in both open-source and industrial test suites. 

On the other end of the spectrum, three smells appear in the majority of the projects: the symptom \emph{Hard Coded Values} appears in more than 90\% of the tests, \emph{Over Checking} between 75\% (industrial) and 79.5\% (open-source), and finally \emph{Sneaky Checking} appears in 70\% of the tests.

Finally, an interesting observation concerns the presence of significant difference between industrial and open-source projects. Notably, in industrial projects three symptoms appear much more often than in open-source projects, namely, \emph{Middle Man} (93.5\%), \emph{Army of Clones} (91.2\%) and \emph{Long Test Steps} (58.8\%). The results for the \emph{Army of Clones} and \emph{Middle Man} can be explained by the structure of the projects at BGL BNP Paribas. Symptoms for the SUIT smell \emph{Army of Clones} manifest because many repositories are interconnected, leading to code duplication across projects.
Hence, testers working on different applications or different parts of the system test the same functionality over and over. As for the symptoms of \emph{Middle Man}, their presence is explained by the use of \emph{Keywords} that act as translation layers. While two actions can perform the same concrete actions on the SUT, in different contexts they might operate on a different business logic.
Thus, developers created translation layers to unify the vocabulary used in the acceptance layer of each test. These translation layers are implemented by the use of a \emph{Keyword} only calling one other \emph{Keyword} with a different name. Finally, \emph{Long Test Step} originates from the complex business logic that is expressed by the step in a test at BGL BNP Paribas and is more related to the SUT and the vocabulary employed by the business analysts than the test suite itself.

\begin{figure}
\centering
\includegraphics[width=0.9\linewidth]{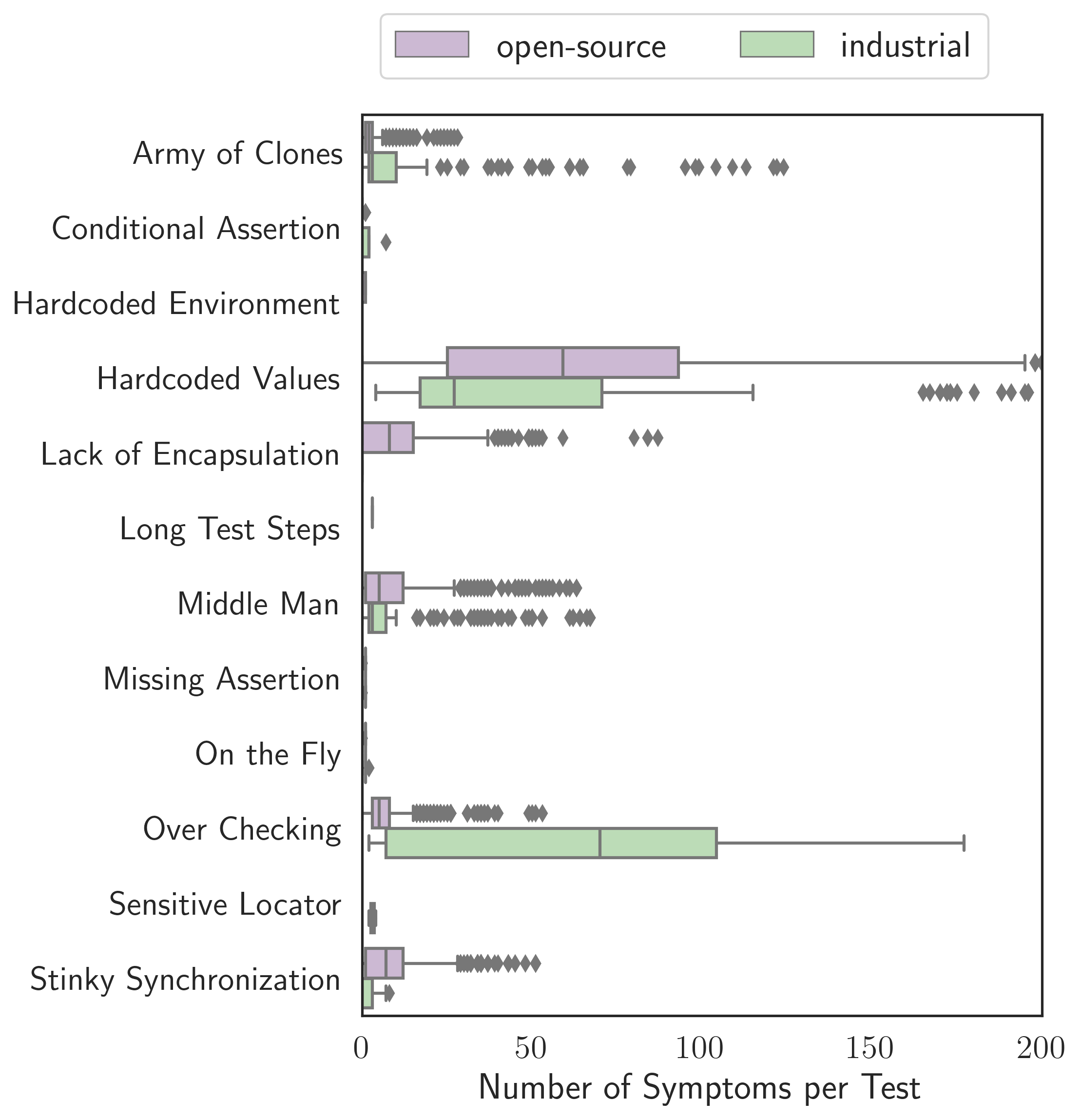}
\caption{Distribution of the number of symptoms in symptomatic tests for all projects across all commits.}  
\label{fig:raw-smells-distribution}
\end{figure}

We continue our analysis with the evaluation of the number of symptoms appearing per tests (Figure~\ref{fig:raw-smells-distribution}). Note that only tests presenting at least one symptom are considered in our analysis, \emph{i.e.} symptomatic tests. Looking at \emph{Hard Coded Values}, we see that even both industrial and open-source projects exhibit the same proportion of tests affected by \emph{Hard Coded Values}, the number varies significantly from a median of 27 in the case of industrial projects to 59 for open-source ones. A similar observation can be made for the case of \emph{Over Checking} where the median varies from 5 (open-source) to 70 (industrial). Finally, the last significant difference regards the symptom \emph{Army of Clones}. Not only the number of tests containing duplicated code (Figure~\ref{fig:diffusion}) is higher, but the average number of duplicated \emph{Keywords} in symptomatic tests is higher in industrial projects (3) than in open-source (2).


\begin{figure}
\centering
\includegraphics[width=0.9\linewidth]{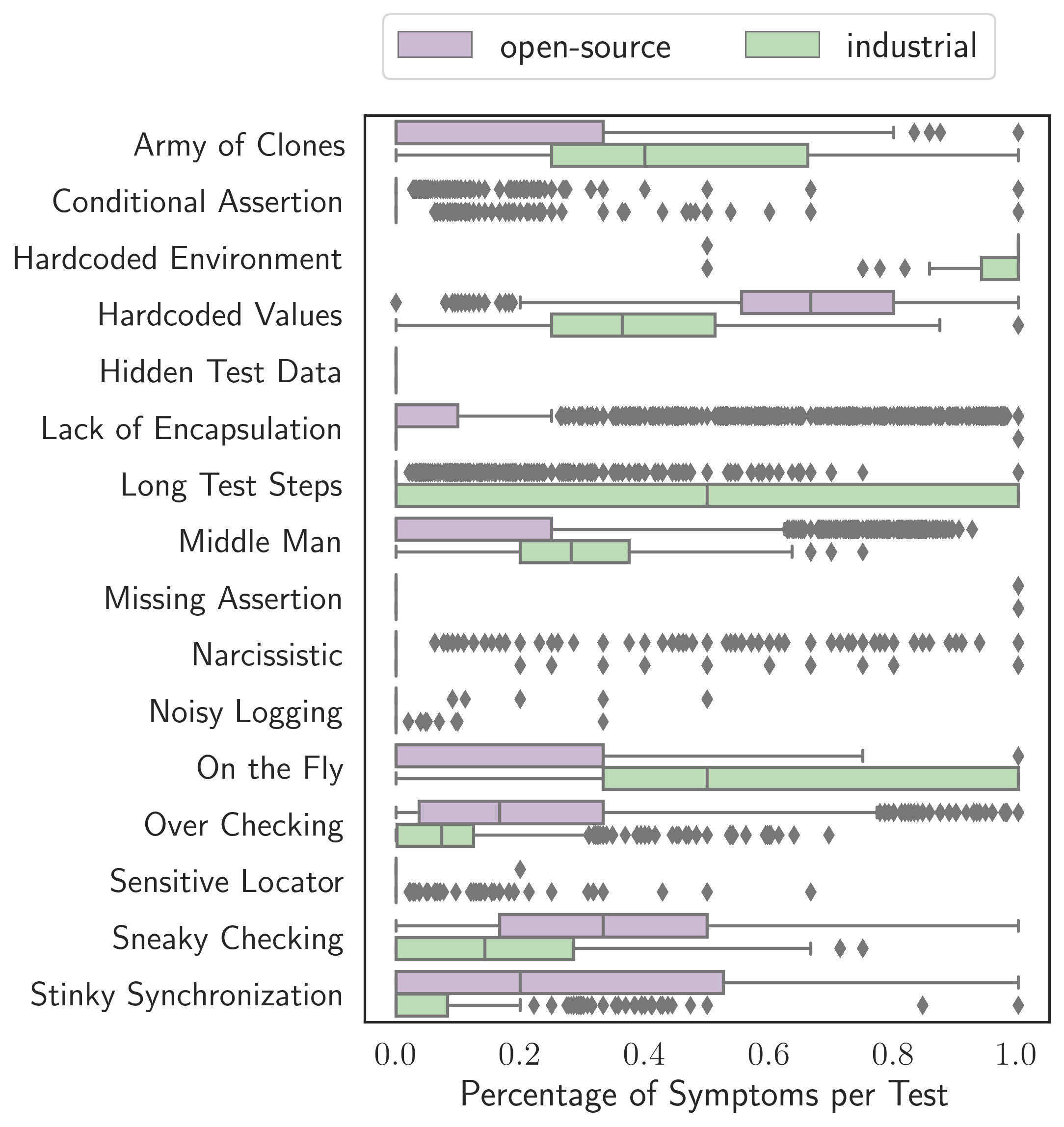}
\caption{Distribution of the density of symptoms in each test for all projects across all commits where the density is the percentage of locations that a symptom could appear at which actually exhibit the symptom.}  
\label{fig:normalized-smells-distribution}
\end{figure}

To put the results of Figure~\ref{fig:raw-smells-distribution} into perspective, we show the density metrics, $D_s(t)$, in Figure~\ref{fig:normalized-smells-distribution}. As a reminder, the density value presents the number of symptoms appearing in a SUIT divided by the number of times it could have appeared. One interesting finding to put in contrast with the previous figure regards the smell \emph{Hardcoded Environment}. While the symptoms do not appear often (only outliers present a value greater than 0 in Figure~\ref{fig:raw-smells-distribution}), whenever an environment variable is used, it will generally be through the use of a hardcoded value which is shown by a median value for its density metric at 1 for both open-source and industrial projects. That is, with regard to its potential occurrences, the SUIT smell \emph{Hardcoded Environment} is actually very frequent.

On the other hand, the smells \emph{Noisy Logging}, \emph{Hidden Test Data}, \emph{Sensitive Locator}, \emph{Narcissistic} and \emph{Conditional Assertion} present low scores in the density value. This means that even when the conditions for their occurrence are present, the symptoms  of these smells do not manifest.


To assess its potential to appear in a test suite, we consider a symptom as frequent if the median value of the density metric is greater than 0. Thus, \emph{Hardcoded Values} with median values of 0.67 and 0.33, \emph{Over Checking} with median values of 0.17 and 0.08, and \emph{Sneaky Checking} with median values of 0.33 and 0.14, for open-source and industrial projects respectively, are frequent in both industrial and open-source projects. This aligns with the observations from  the count metric and confirms the prevalence of these smells. Similarly, following the same trend from Figure~\ref{fig:raw-smells-distribution} \emph{Army of Clones} (median value of 0.40), \emph{Long Test Step} (median value of 0.5) and \emph{Middle Man} (median value of 0.29) appear often in industrial projects.

Nonetheless, in the case of the industrial project, we observe that \emph{On the Fly} with a median value of 0.67 shows a deviation from the count metric, $S_s(t)$. 
As for open-source projects, we observe that \emph{Sticky Synchronization} is frequent, with a median value of 0.2. The divergence with the count metric can be explained by the fact that many tests do not use explicit synchronization mechanism, thus, limiting the number of tests where the symptom could appear. Furthermore, the difference from the industrial project can explained by the hard policy in the QA team at BGL BNP Paribas, where explicit timeouts are considered as a major source of flakiness. Indeed, explicit timeouts are strongly discouraged by the QA team and are usually removed during code reviews.

Both industrial and open-source projects present the some similar trends despite their difference in terms of design and architecture. We observe that 7 out of the 16 smells symptoms do not typically appear in SUITs, whereas the smells \emph{Hardcoded Values}, \emph{Over Checking}, and \emph{Sneaky Checking} are frequent. As for the difference observed between industrial and open-source projects, we note that the smells \emph{Army of Clones}, \emph{Long Test Step}, \emph{Middle Man} and \emph{On the Fly} often exhibit symptoms in the industrial project but not in the open-source projects and inversely symptoms of \emph{Sticky Synchronization} tend to appear more often in open-source projects. Finally, the symptoms of \emph{Hardcoded Environment} appearing systematically would suggest that the developers do not consider it as a bad practice.

Finally, we perform a ranking analysis between industrial projects and open-source projects. The goal of this analysis is to determine if the symptoms appear in the same order in both project. Thus, we use the normalized Levenshtein similarity where the order of the symptoms is determined by the mean number of symptoms and percentage of symptom. This metric provide an indication as the number of permutations that are necessary in order to go from one list to the other, in other words their similarity. A value of $1$ indicates a perfect similarity and a value of $0$ that the two list are total dissimilar. We obtain a value of $0.1819$ when comparing the number of symptoms and a value of $0.3125$ in the case of the number of symptoms. These low scores suggest that the relative importance of the symptoms present in SUITs differs from industrial project and open-source projects.

\begin{tcolorbox}[colframe=black!50!white, arc=10pt,before skip=10pt plus 2pt,after skip=10pt plus 2pt]
    The smells \emph{Hard Coded Values}, \emph{Over Checking}, and \emph{Sneaky Checking} are prevalent in both industrial and open-source projects. These smells appear in large numbers and exhibit a high density compared to their potential number of occurrences.
    Due to project and team structure, the industrial project is more prone to complexity smells such as \emph{Army of Clones} and \emph{Middle Man}.
    On the other hand, their code reviewing policies seem to prevent the symptoms of \emph{Stinky synchronization}.
\end{tcolorbox}


\subsection{RQ3: SUIT Smells Refactoring}
\label{sec:results-smell-refactoring}

\begin{table}
\centering

\begin{tabularx}{0.85\textwidth}{>{\raggedright}p{1.3in}>{\raggedleft}p{0.5in}>{\raggedleft}p{0.5in}>{\raggedleft}p{0.5in}>{\raggedleft}p{0.5in}}

\hline
& \multicolumn{2}{c}{\textbf{Industrial}} & \multicolumn{2}{c}{\textbf{Open-source}} \tabularnewline
\textbf{Symptom} & \textbf{Count} & \textbf{Percent}  & \textbf{Count} & \textbf{Percent} \tabularnewline
\hline
Army of Clones & 738 & 36.64 & 10,139 & 24.42 \tabularnewline
Conditional Assertions & 9 & 4.95 & 270 & 0.32 \tabularnewline
Hardcoded Environment & 0 & 0.00 & 882 & 9.95 \tabularnewline
Hardcoded Values & 226 & 18.44 & 28,863 & 17.09 \tabularnewline
Hidden Test Data & 0 & 0.00 & 0 & 0.00 \tabularnewline
Lack of Encapsulation & 8 & 0.00 & 10,944 & 11.52 \tabularnewline
Long Test Steps & 0 & 0.00 & 34 & 0.19 \tabularnewline
Middle Man & 1,037 & 39.57 & 55,509 & 27.62 \tabularnewline
\textbf{Missing Assertion} & \textbf{6,647} & \textbf{90.45} & \textbf{137,707} & \textbf{72.86} \tabularnewline
Narcissistic & 0 & 0.00 & 0 & 0.00 \tabularnewline
Noisy Logging & 0 & 0.00 & 0 & 0.00 \tabularnewline
On the Fly & 27 & 13.85 & 516 & 7.93 \tabularnewline
Over-Checking & 35 & 3.46 & 21,586 & 15.50 \tabularnewline
Sensitive Locators & 2 & 4.55 & 0 & 0.00 \tabularnewline
Sneaky Checking & 0 & 0.00 & 0 & 0.00 \tabularnewline
Stinky Synchronization & 38 & 4.92 & 23,163 & 23.28 \tabularnewline
\hline
\end{tabularx}

\caption{Total Number of refactoring actions (\emph{Count}) and percentage of symptomatic tests where at least one refactoring action was performed during their lifetime (\emph{Percent}) for industrial and open-source projects.}
\label{tab:fixes}
\end{table}

While Section~\ref{sec:results-smells-diffusion} focuses on the prevalence of smell symptoms across SUITs, this section tackles the question of how often the symptoms are refactored by practitioners. We use the methodology described in Section~\ref{sec:methodology-smell-evolution} to extract the fine-grained changes removing a symptom from the test, \emph{i.e.} refactoring actions.

Column \emph{Count} in Table~\ref{tab:fixes} shows the number of refactoring actions across all SUIT-modifying commits. Adding assertions to tests presenting the symptom \emph{Missing Assertion} is the most common type of refactoring action as it occurs 6,647 times in the industrial project and 137,707 times in open-source projects. This result is explained by the fact that in SUITs, creating a scenario and being able to run it from beginning to end already provide a signal. However, once the test is ready and behaves as expected, more specific checks, \emph{i.e.} assertions, are added to improve its readability and its fault detection capabilities. Thus, we observe that some SUITs are missing assertion when created but assertions are added in later commits. 

In the industrial project, three other types of refactoring actions see high values compared to the rest, namely, \emph{Army of Clones} with 738 refactoring actions, \emph{Hardcoded Value} with 226 refactoring actions, and \emph{Middle Man} with 1,037 refactoring actions. Indeed, with the introduction of the tool presented in \cite{Rwemalika2019b}, the team at BGL BNP Paribas became aware of the existence of a large amount of code duplication and actively started to work on reducing it, which explains the number of refactorings for \emph{Army of Clones}. As for \emph{Middle Man}, during the year 2019, the team performed a normalization in the naming of \emph{Keywords}, in an effort to improve readability in the codebase. Consequently, names such as ``Fill Form Next Page'' were changed to more expressive forms such as ``Fill Login Form and Validate''. The goal was to increase the expressiveness of the test codebase and consequently, it reduced the need for a translation layer. Note that in this case, the team was targeting another SUIT smell, \emph{Unsuitable Naming}, where the name of the \emph{Keyword} does not provide indication as what it is doing, but ended up also addressing another smell, \emph{Middle Man}. Finally, observing the number of refactoring actions addressing \emph{Hardcoded Value} is mainly due to the fact that the symptom appears often. Indeed, when observing the column \emph{Percent} from Table~\ref{tab:fixes}, we see only 18\% of the test exhibiting the symptom are refactored through their lifetime. 

Still focusing on the results for the industrial project, for some symptoms, we never observe refactoring actions. This is the case for: \emph{Hidden Test Data}, \emph{Noisy Logging}, \emph{Narcissistic}, \emph{Sensitive Locator} and \emph{Sneaky Checking}. With the exception of \emph{Sneaky Checking}, these symptoms only rarely occur in the test codebase. Hence, in the absence of symptoms, there is nothing to refactor. \emph{Sneaky Checking}, on the other hand, is considered frequent according to our metric, but is never addressed. One reason can be that to increase readability, assertions in the acceptance criteria are often encapsulated in a \emph{User Keyword} with more meaningful names. When looking at the metric \emph{Narcissistic} for the industrial project, we observe no fine-grained refactoring. 

In Listing~\ref{fig:robot-script} (Section~\ref{sec:background-keyword-driven-testing}) we present an example of a KDT extracted from the official document of Robot Framework. We observe at line 9 a call to ``welcome page should be open'' which exhibits the symptom for \emph{Sneaky Checking}. However, calling directly the \emph{Library Keyword} ``Title Should be'' would have decreased the readability of the test and introduced another symptom, \emph{Lack of Encapsulation}.

\begin{figure}
\centering
\includegraphics[width=0.5\linewidth]{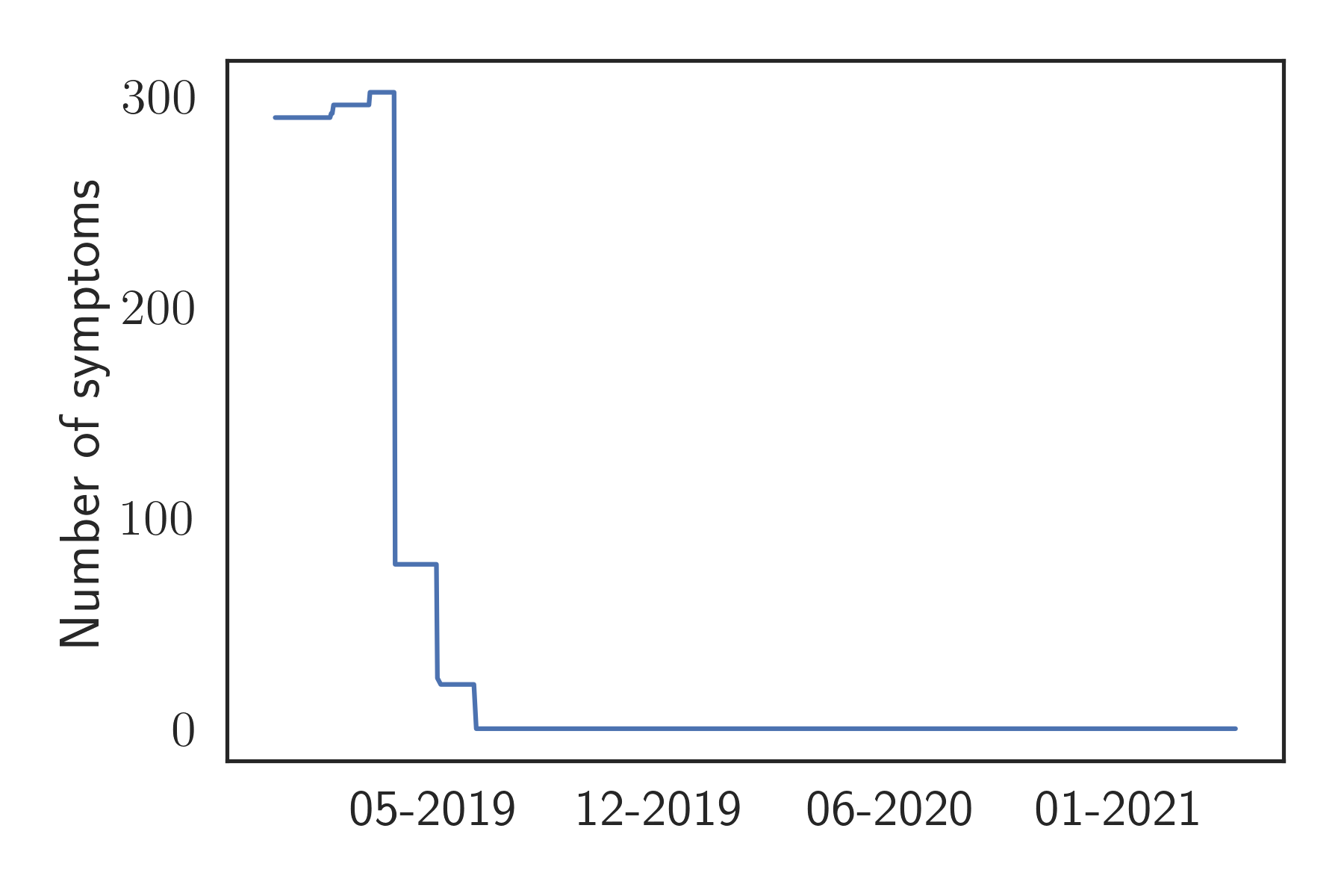}
\caption{Evolution of the number of symptoms for the smell \emph{Narcissistic} over time for the industrial project (BGL BNP Paribas).}  
\label{fig:evolution-bgl-narcissistic}
\end{figure}

Putting these results into perspective, Figure~\ref{fig:evolution-bgl-narcissistic} shows the evolution of the symptom \emph{Narcissistic} over time. We observe an abrupt decrease of the number of symptoms until they completely disappear. Where this observation seem to contrast with previous results, this is explained by old tests presenting this pattern are deprecated by the team and replaced by new ones not presenting the smell. Thus, while there is no specific refactoring action happening, the symptoms are removed from the test suite as new test are introduced and old tests deprecated. 

The results for open-source projects depict a similar picture. Symptoms for \emph{Missing Assertion}, \emph{Middle Man}, \emph{Army of Clones}, and \emph{HardCoded Values} show also relatively high values compared to other symptoms. Similarly, we also notice a relatively large number of refactoring actions for the symptoms of \emph{Stinky Synchronization}, \emph{Over Checking}, and \emph{Lack of Encapsulation}. The difference between the results from \emph{Stinky Synchronization} can explain by the active policy in the industrial team preventing such symptoms to appear altogether.

Finally, similarly to Section~\ref{sec:results-smells-diffusion}, we perform a rank analysis using the normalized Levenshtein similarity. The value obtained when comparing the number of fixes performed in open-source and industrial projects is $0.4375$ and $0.2667$ when accounting for the percentage of refactoring. These low values show that the majority of refactoring operations are not ranked in the same way in both industrial and open-source projects. Taking into account the results of Section~\ref{sec:results-smells-diffusion} this comparison suggests that when generalizing results obtained from open-source projects, researchers should remain careful to their applicability in industrial contexts.

In conclusion, the results presented in this section show that the refactoring operations performed in industry and open-source projects are different in type and frequency. However, in both cases, the proportion of symptomatic tests that are refactored remain low with the exception of the symptoms for \emph{Missing Assertion}. 

\begin{tcolorbox}[colframe=black!50!white, arc=10pt,before skip=10pt plus 2pt,after skip=10pt plus 2pt]
    For half of the smells less than 50\% of the symptomatic tests are ever subjects of refactoring. \emph{Missing assertion} is a unique exception with between 70\% (open-source) and 90\% (industrial) of the symptomatic tests being refactored. Interestingly, while refactoring actions are rare, test smells like \emph{Narcissistic} and \emph{Middle Man} still disappear from the test codebase as a side effect of the replacement of symptomatic tests by new tests not exhibiting the symptom.
\end{tcolorbox}

\section{Threats to Validity}

Threat to construct validity result from the non suitability of the metrics used to evaluate the results. To detect test smells we rely on heuristics based on code metrics. While we focus on metric that offer good precision, we cannot ignore the fact that some of the smells are not detected by the metrics devised in this work.

Threats to the internal validity are due to the design of the study, potentially impacting our conclusions. Such threats typically do not affect exploratory studies like the one in this work. A caveat can be raised on how the changes are extracted. Indeed, changes are recorded when developers check-in their changes to the control version system. Thus, refactoring actions might be lost if developers do not check-in often their changes or on the contrary, we might flag artifacts of the development process (\emph{e.g.} Assertions added only at the end of the test) as refactoring actions. To account for this phenomenon, we analyze manually a subset of the results to ensure the soundness of the process.

Finally, the threats to external validity, regarding the generalization of the results, concerns mainly the choice of the projects analyzed. Indeed, conducting our analysis on a limited sample of projects, our results might not generalize to other projects. However, we try to control for this limitation by selecting projects of different sizes, from different domains and in different development cultures. Furthermore, working with Robot Framework, there is no guarantee that the results presented in this work are transferable to other languages or technologies.

\section{Related work}

Considering the uniqueness and impacts of test smells, the research community entailed studies to extend the catalogs of known test smells, empirical studies have been conducted to analyze their introduction, diffusion and suppression, and finally, tools have been proposed to automatically detect them.

\cite{VanDeursen2001} are among the first to introduce the concept of test smells. They describe a catalog of 11 test smells and refactoring operation to address them. Following their work, different studies were conducted to further extend the catalog of test smells. Departing from the analysis of the code to identify test smells, \cite{Bowes2017} and \cite{Tufano2016} conducted human studies to isolate test smells and report the perceived impacts on both test code and production code.

With good test smell catalogs established, the community investigated the diffusion and evolution of smells in test code. \cite{Bavota2015} conducted an empirical studies to analyze the diffusion and the impact of test smells. In their work, \cite{Tufano2016} and later \cite{Kim2020} showed that test smells are introduced when the test is written and are long lived.

To help practitioners identify potential sub-optimal patterns in their test codebase, researchers introduced tooling to automatically detect them. Typically, detection tools analyze test code metrics using rule based heuristics \citep{VanRompaey2007, Reichhart2007, Peruma2020} to isolate test smells.

Nonetheless, all of the work mentioned above focuses on test smells present in unit testing. \cite{Hauptmann2013} conducted a study on test smells present in tests expressed in natural language in industrial systems. However, their experiments show that the metrics extracted cannot be used for assessment of the quality of a test suite. \cite{Femmer2014, Femmer2017} introduce the concept of requirement smells and conducted an empirical evaluation of their approach using industrial projects. In their work, the author present an automated static analysis technique relying on natural language processing (NLP) to detect smells in the requirements. The main goal at the requirement level is to detect whether or not a requirement contains ambiguities, therefore, the techniques developed rely on concrete instances of ambiguities. Finally \cite{Chen2012}, propose a catalog of 11 test smell present in KDT along with 16 refactoring methods.

\section{Conclusion and Implications}

The goal of this paper is to shed light on the smells occurring in SUITs. To do so, we conducted  a multivocal literature review and identified SUIT-specific smells from both formal and grey literature. This process lead to a catalog of 35 SUIT smells. For 16 of these smells we derive metrics to characterize their presence and refactoring in the test code. Our results show that three smells \emph{Hardcoded Values} (90\% of the tests), \emph{Over Checking} (between 75\% and 80\% of the tests), and \emph{Sneaky Checking} (70\% of the tests) are prevalent in both industrial and open-source projects. 
While these symptoms are largely present in the test codebase, their refactoring on the other hand remains low with only less than half the tests ever observing any targeted refactoring action with the exception of \emph{Missing Assertion}.

Contrastingly, one symptom from the catalog never appears in the subjects of our study, namely, \emph{Hidden Test Data}. Similarly, other symptoms tend to only appear in rare occasions such as \emph{Sensitive Locator}, \emph{Noisy Logging} and \emph{Conditional Assertion}. Finally, the smell \emph{Hardcoded Environment} while remaining rare, appears systematically when an environment variable is required.

Though we observe general trends common to both industrial and open-source projects, when performing a more rigorous comparison between the two sets of projects, we observe significant differences. Indeed, the projects differ both in terms of prevalence of the symptoms and refactoring operations. This results can be explain by the difference in scope, actors, and lifecycle present in each context. Consequently, these results suggest that when conducting studies researchers should be aware of these fundamental differences which might limit the generalization of their results.

In light of the results from this exploratory analysis, we believe that extending the catalog of known SUIT smells will have an impact on our partner's development process. Indeed, they already started using our tooling to address some of the bad design choices which we observed in their test codebase. Moreover, with our new catalog and these first observations, we open perspective for future research on awareness of bad testing practices and the pitfalls to avoid when evolving SUITs. 

This study alongside with the tooling developed and the available dataset\footnote{https://github.com/kabinja/suit-smells-replication-package} lays the ground for future research on the impact of smells on SUIT suites. Our future agenda is focused on devising automated refactoring approaches. We also plan to extend the list of 16 metrics to cover more smells from our catalog. 




\bibliographystyle{spbasic} 
\bibliography{cost-of-smell}

\end{document}